\documentclass{article}

\usepackage{arxivsimplified}

\usepackage[utf8]{inputenc} 
\usepackage[T1]{fontenc}    
\usepackage{hyperref}       
\usepackage{url}            
\usepackage{booktabs}       
\usepackage{amsfonts}       
\usepackage{nicefrac}       
\usepackage{microtype}      
\usepackage{lipsum}

\usepackage{graphicx}
\graphicspath{{figures_pdf/}}
\usepackage{subfigure}
\usepackage{cite}
\usepackage[dvipsnames]{xcolor}
\usepackage{amsmath}
\usepackage{blindtext}
\usepackage{amsfonts}
\usepackage{multirow}
\usepackage{array}
\usepackage{mathtools}
\usepackage{listings}
\usepackage{xcolor}
\definecolor{codegreen}{rgb}{0,0.6,0}
\definecolor{codegray}{rgb}{0.5,0.5,0.5}
\definecolor{codepurple}{rgb}{0.58,0,0.82}
\definecolor{backcolour}{rgb}{0.95,0.95,0.92}
\usepackage{esvect}
\usepackage{tabularx}
\newcolumntype{L}[1]{>{\raggedright\arraybackslash}p{#1}}
\newcolumntype{C}[1]{>{\centering\arraybackslash}p{#1}}
\newcolumntype{R}[1]{>{\raggedleft\arraybackslash}p{#1}}
\newcommand{\overbar}[1]{\mkern 1.5mu\overline{\mkern-1.5mu#1\mkern-1.5mu}\mkern 1.5mu}

\usepackage{algorithm}
\usepackage{algorithmic}
\usepackage{enumitem}
\setlist[itemize]{leftmargin=*}

\lstdefinestyle{mystyle}{
    backgroundcolor=\color{backcolour},   
    commentstyle=\color{codegreen},
    keywordstyle=\color{magenta},
    numberstyle=\tiny\color{codegray},
    stringstyle=\color{codepurple},
    basicstyle=\ttfamily\footnotesize,
    breakatwhitespace=false,         
    breaklines=true,                 
    captionpos=b,                    
    keepspaces=true,                 
    numbers=left,                    
    numbersep=5pt,                  
    showspaces=false,                
    showstringspaces=false,
    showtabs=false,                  
    tabsize=2
}
 
\usepackage{scalerel,stackengine}
\stackMath
\newcommand\reallywidehat[1]{%
\savestack{\tmpbox}{\stretchto{%
  \scaleto{%
    \scalerel*[\widthof{\ensuremath{#1}}]{\kern-.6pt\bigwedge\kern-.6pt}%
    {\rule[-\textheight/2]{1ex}{\textheight}}
  }{\textheight}%
}{0.5ex}}%
\stackon[1pt]{#1}{\tmpbox}%
}

\title{Long short-term memory embedded nudging schemes for nonlinear data assimilation of geophysical flows}

\author{
  Suraj Pawar  \\
  School of Mechanical \& Aerospace Engineering,\\
  Oklahoma State University, \\
  Stillwater, Oklahoma - 74078, USA.\\
  \texttt{supawar@okstate.edu} \\
   \And
   Shady E. Ahmed  \\
  School of Mechanical \& Aerospace Engineering,\\
  Oklahoma State University, \\
  Stillwater, Oklahoma - 74078, USA.\\
  \texttt{shady.ahmed@okstate.edu } \\
  \And
 Omer San \\
 School of Mechanical \& Aerospace Engineering,\\
  Oklahoma State University, \\
  Stillwater, Oklahoma - 74078, USA.\\
  \texttt{osan@okstate.edu} \\
  \And
  Adil Rasheed  \\
  Department of Engineering Cybernetics,\\
  Norwegian University of Science and Technology,\\
  N-7465, Trondheim, Norway.\\
  \texttt{adil.rasheed@ntnu.no}\\
  \And
  Ionel M. Navon  \\
  Department of Scientific Computing,\\
  Florida State University,\\
  Tallahassee Florida - 32306, USA.\\
  \texttt{inavon@fsu.edu}\\
}

\begin{document}
\maketitle

\begin{abstract}
Reduced rank nonlinear filters are increasingly utilized in data assimilation of geophysical flows, but often require a set of ensemble forward simulations to estimate forecast covariance. On the other hand, predictor-corrector type nudging approaches are still attractive due to their simplicity of implementation when more complex methods need to be avoided. However, optimal estimate of nudging gain matrix might be cumbersome. In this paper, we put forth a fully nonintrusive recurrent neural network approach based on a long short-term memory (LSTM) embedding architecture to estimate the nudging term, which plays a role not only to force the state trajectories to the observations but also acts as a stabilizer. Furthermore, our approach relies on the power of archival data and the trained model can be retrained effectively due to power of transfer learning in any neural network applications. In order to verify the feasibility of the proposed approach, we perform twin experiments using Lorenz 96 system. Our results demonstrate that the proposed LSTM nudging approach yields more accurate estimates than both extended Kalman filter (EKF) and ensemble Kalman filter (EnKF) when only sparse observations are available. With the availability of emerging AI-friendly and modular hardware technologies and heterogeneous computing platforms, we articulate that our simplistic nudging framework turns out to be computationally more efficient than either the EKF or EnKF approaches.  
\end{abstract}


\section{Introduction}
\label{sec:intro}
Data assimilation (DA) is a methodology where the observations are utilized to correct the results from a mathematical model to reconstruct spatiotemporal dynamics of a system \cite{lewis2006dynamic,simon2006optimal,evensen2009data}.
DA is used extensively for weather forecasting, where there is a growing number of observations coming from satellites, and in situ monitoring. Variational and sequential schemes are two of the most widely used approaches in dynamical data assimilation. For the former, DA is formulated as a minimization problem, where the objective function is defined as the discrepancy between real observations and model's predictions based on a given set of initial conditions and parameters. The argument of this minimization problem is the set of model's initial conditions and parameters that need to be tuned to drive the predictions towards the observations. On the other hand, sequential methods usually rely on statistical inference using Bayesian analysis, where the current measurements are used to correct the prior model forecasts to get a better posterior estimate, usually called the analysis in DA terminology. 

One of the key limitations of DA methods is that they rely on a forward model whose dynamics is known. For high-dimensional systems like geophysical flows, standard DA methods suffer from the curse of dimensionality. With the increasing resolution of numerical models, the nonlinearities are likely to become so strong that DA algorithms based on linearization might fail \cite{van2009particle}. In recent years, with an explosion of data generated from observations, experimental measurements, and numerical simulations, there is a growing interest in applying data-driven methods along with DA \cite{lguensat2017analog}. Mostly, efforts focused on using data-driven models in lieu of conventional (physics-based) models in order to accelerate the DA computations. Tang et al.\cite{tang2019deep} developed a surrogate based model based on convolutional and recurrent neural networks for predicting dynamical subsurface flows and employed it in the DA framework as an emulator to the forward dynamical model. There have been several other studies that demonstrated the potential of data-driven methods in the accurate prediction of complex physical systems such as flooding \cite{hu2019rapid}, global atmospheric model \cite{szunyogh2020machine}, quasigeostrophic flows \cite{rahman2019non}, chaotic systems \cite{pathak2018model,vlachas2018data}, soil water dynamics\cite{li2020comparison}, and tsunami modeling \cite{cheng2020data}. Recent works have also drawn ideas to synthesize DA with reduced order models \cite{protas2015optimal,zerfas2019continuous,xiao2018parameterised,daescu2007efficiency,cstefuanescu2015pod,cao2007reduced,robert2005reduced,arcucci2019optimal,brunton2019machine,puzyrev2019pyrom}.
Bocquet et al.\cite{bocquet2020bayesian} proposed a hybrid framework by combining DA and machine learning (ML) to estimate the model, the state trajectory, and model error statistics for high-dimensional chaotic systems from partial and noisy observations. Brajard et al.\cite{brajard2020combining} proposed an algorithm where neural networks provide a surrogate forward model to DA and DA provides a time series of complete states to train the neural network. They illustrated the convergence of proposed algorithm for the Lorenz 96 system and achieved the accurate forecasts \textcolor{black}{up to two Lyapunov time units}. 

Correspondingly, ML tools can also benefit from DA algorithms. Abarbanel et al.\cite{abarbanel2018machine} offer a perspective on the equivalence between ML and statistical data assimilation and discuss how methods developed in DA can be potentially useful for ML. Bocquet et al.\cite{bocquet2019data} proposed DA as a learning tool to infer ordinary differential equations for dynamical systems solely from noisy data and showed its connection with deep learning methods. P{\'e}rez-Ortiz et al.\cite{perez2003kalman} showed that the long short-term memory (LSTM) network can be trained efficiently with better generalization using the decoupled extended Kalman filter \cite{puskorius1994neurocontrol}.

As an extension to the current efforts of using ML tools in DA context, we propose a modular neural-network based DA framework. In other words, we utilize ML to achieve the fusion between the model's estimates and noisy observations to provide more accurate predictions, rather than using ML as a facilitator to just accelerate existing DA algorithms. To accomplish this, we train a long short-term memory (LSTM) neural network to ``nudge'' model's forecast given a set of sparse observations. Nudging is a relatively simple DA approach that uses the forecast error, defined as the difference between model predictions and measurements, to constrain and correct the model evolution. Nudging was introduced by Anthes\cite{anthes1974data} for the initialization of hurricane models from real observational data. In nudging methods, the state analysis is approximated as a linear superposition between its model forecast and forecast error. Despite its conceptual simplicity, nudging schemes often require adhoc approximation of the nudging (or weighting) matrix. In our framework, we relax this linear superposition assumption and avoid those adhoc approximations by training an LSTM neural network to \emph{nonlinearly} blend model's forecast and sparse observations.

We demonstrate and test the proposed LSTM-DA framework using the Lorenz 96 system as a benchmark problem in geophysical science applications. We illustrate the success of LSTM-DA using different sets of observations with varying levels of noise and sparsity. In particular, we consider combinations between data-rich, data-deficient, observation-rich, and observation-deficient settings. We also compare our results against some of the common DA techniques. Namely, we discuss the results of extended Kalman filter (EKF), ensemble Kalman filter (EnKF), deterministic ensemble Kalman filter (DEnKF), and a simple forward nudging method. Our LSTM-DA framework can be considered very much similar to the methodology proposed by Zhu et al.\cite{zhu2019model} in which the fully connected neural network was used to learn the uncertainty in the mathematical model arising from linearization, discretization, and model reduction. The difference in our proposed framework is that we employ the LSTM neural network to learn the nudging correction term in order to cure the discrepancy between prior predictions and measurements that might arise due to inaccurate initial conditions, boundary conditions, or model parameters.

The rest of the manuscript is outlined here. In Section~\ref{sec:kalman}, we describe three of the most common nonlinear filtering techniques as benchmarks to compare our framework against. In particular, we briefly outline the extended Kalman filter, which is a first-order adaptation of the standard Kalman filter to deal with nonlinear models. We then introduce the ensemble Kalman filter and its deterministic version as reduced rank variants of nonlinear filters. In Section~\ref{sec:nudging}, we discuss the nudging method as a simple alternative to nonlinear filters, which is then extended as a base for our proposed DA-LSTM framework in Section~\ref{sec:LSTM}. We define the DA set-up using Lorenz 96 system in Section~\ref{sec:L96}. After that, we provide our results in Section~\ref{sec:results} as well as relevant discussions and comparisons using different sets of historical data and observations. Finally, we draw our conclusions as well as the limitations and potential extensions of the present study in Section~\ref{sec:conc}.









\section{Nonlinear filtering}
\label{sec:kalman}
The central goal of DA is to extract the information from observational data to correct dynamical models and improve their prediction. There are different approaches such as variational methods like 4D-Var and stochastic methods like ensemble filters that are widely used in DA. Several textbooks on data assimilation offer academic explanations and discussion on these methods \cite{lewis2006dynamic,evensen2009data,simon2006optimal,kalnay2003atmospheric}.

In this section, we discuss sequential data assimilation problem and then outline the algorithm procedure for extended Kalman filter (EKF) and ensemble Kalman filter (EnKF). The complete derivation of Kalman filter and its different variants can be found in a number of literature \cite{lewis2006dynamic,simon2006optimal,evensen2009data,gelb1974applied,welch1995introduction}.

For demonstration, we consider the dynamical system whose evolution is governed by 
\begin{equation}\label{eq:dyn_model}
    \mathbf{x}_{k+1} = \textbf{M}(\mathbf{x}_{k}) + \mathbf{w}_{k+1},
\end{equation}
where $\mathbf{x}_k \in \mathbb{R}^n$ is the state of the dynamical system at discrete time $t_k$, $\textbf{M}:\mathbb{R}^n \rightarrow \mathbb{R}^n$ is the nonlinear model operator that defines the temporal evolution of the system. The term $\mathbf{w}_{k+1}$ denotes the model noise that takes into account the mathematical model error, numerical approximations, and the boundary conditions. In our study we assume that the model noise is drawn from a multivariate normal distribution with zero mean and a covariance matrix $\mathbf{Q}_k$, i.e., $\mathbf{w}_k \sim {\cal{N}}(0,\mathbf{Q}_k)$. 

Let $\mathbf{z}_k \in \mathbb{R}^m$ be observations of the state vector obtained through noisy measurements procedure as given below
\begin{equation}
    \mathbf{z}_k = h(\mathbf{x}_k) + \mathbf{v}_k,
\end{equation}
where $h(\cdot)$ is a nonlinear function that maps $\mathbb{R}^n \rightarrow \mathbb{R}^m$, also known as the observational operator defining a map between state space and measurement space, and $\mathbf{v}_k \in \mathbb{R}^m$ is the measurement noise. We assume that the measurement noise is a white Gaussian noise with zero mean and the covariance matrix $\mathbf{R}_k$, i.e., $\mathbf{v}_k \sim {\cal{N}}(0,\mathbf{R}_k)$. Furthermore, we assume that the noise vectors $\mathbf{w}_{k}$ and $\mathbf{v}_{k}$ at two different time steps are uncorrelated, which is a common assumption in sequential data assimilation problems. In sequential data assimilation problems, the objective is to estimate the state $\mathbf{x}_k$ given the observations up to time $t_k$, i.e., $\mathbf{z}_1,\dots,\mathbf{z}_k$. When we use observations to estimate the state of the system, we say that the data are assimilated into the model. There is a number of studies that deal with non-Gaussian distributions for noise vectors \cite{li2018trimmed,anderson2010non,apte2007sampling}. However, this is outside the scope of this study and we restrict to assumption of Gaussian noise for model and measurement errors. 

We will use the notation $\widehat{\mathbf{x}}_k$ to denote an analyzed state of the system at time $t_k$ when all of the observations up to and including time $t_k$ are used in determining the state of the system. When all the observations before (but not including) time $t_k$ are utilized for estimating the state of the system, then we call it the forecast estimate and denote it as $\mathbf{x}^f_k$. We use the notation $\mathbf{P}_k$ to denote the error covariance matrix. The error covariance matrix for the state vector $\mathbf{x}_k$ is defined as 
\begin{equation}
    \mathbf{P}_k = \text{E}[(\mathbf{x}_k - \text{E}[\mathbf{x}_k])(\mathbf{x}_k - \text{E}[\mathbf{x}_k])^\text{T}],
\end{equation}
where $\text{E}[\cdot]$ denotes the expected value. We use $\widehat{\mathbf{P}}_k$ to denote the error covariance for an analyzed state $\widehat{\mathbf{x}}_k$ and $\mathbf{P}^f_k$ denotes the error covariance for the forecast estimate $\mathbf{x}^f_k$.  

\begin{algorithm}[H]
\caption{Extended Kalman filter}
\label{alg:ekf}
\begin{algorithmic}[1]
\STATE Initialize the state of the system and error covariance. \\
\begin{align}
    \mathbf{\widehat{x}}_0 &= E[\mathbf{x}_0], \label{eq:ekf_x0}\\
    \widehat{\mathbf{P}}_0 &= \mathbf{P}_0. \label{eq:ekf_p0}
\end{align}
\STATE For $k=0,1,\dots$ proceed as follow
\begin{itemize}
    \item Forecast step: Integrate the state estimate and its error covariance from time $t_k$ to $t_{k+1}$ as follow 
    \begin{align}
        \mathbf{x}^f_{k+1} &= \mathbf{M}(\widehat{\mathbf{x}}_k), \label{eq:ekf_xk}\\
        \mathbf{P}^f_{k+1} &= \mathbf{D_M} \widehat{\mathbf{P}}_k \mathbf{D}^\text{T}_\mathbf{M} + \mathbf{Q}_{k+1} \label{eq:ekf_pk},
    \end{align}
    \item Data assimilation step: Once the observations are available at time $t_{k+1}$, they are incorporated into the state estimate and error covariance estimation as follow
    \begin{align}
        \widehat{\mathbf{x}}_{k+1} &= \mathbf{x}^f_{k+1} + \mathbf{K}[\mathbf{z}_{k+1} - h(\mathbf{x}^f_{k+1})], \label{eq:ekf_xa}\\
        \mathbf{K} &= \mathbf{P}^f_{k+1} \mathbf{D}^\text{T}_\mathbf{h}[ \mathbf{D}_\mathbf{h} \mathbf{P}^f_{k+1} \mathbf{D}^\text{T}_\mathbf{h} + \mathbf{R}_{k+1}]^{-1}, \label{eq:ekf_k}\\
        \widehat{\mathbf{P}}_{k+1} &= (\mathbf{I} - \mathbf{K} \mathbf{D_h})\mathbf{P}^f_{k+1} \label{eq:ekf_pa}.
    \end{align}
\end{itemize}
\end{algorithmic}
\end{algorithm} 

\subsection{Extended Kalman filter} \label{sec:ekf}

We first outline the algorithm for extended Kalman filter (EKF) and then discuss in detail its important steps\cite{lewis2006dynamic}. The procedure for the EKF is summarized in Algorithm~\ref{alg:ekf}.

To start with an EKF algorithm, we initialize the state of the system using Equation~\ref{eq:ekf_x0} and error covariance matrix with Equation~\ref{eq:ekf_p0}.
We evolve the state of the system between two observation points (from time $t_k$ to $t_{k+1}$) using the known nonlinear dynamics as given in Equation~\ref{eq:ekf_xk}. The error covariance matrix is propagated between two observation points using Equation~\ref{eq:ekf_pk}. Here $\mathbf{D_M} \in \mathbb{R}^{n \times n}$ is the Jacobain of the model $\mathbf{M}(\cdot)$ and the superscript $\text{T}$ denotes the transpose of the matrix. Once the observation $\mathbf{z}_{k+1}$ becomes available at time $t_{k+1}$, we assimilate it into the forecast state using Equation~\ref{eq:ekf_xa}. The matrix $\mathbf{K} \in \mathbb{R}^{n \times m}$ refers to the Kalman gain matrix and is computed as shown in Equation~\ref{eq:ekf_k}, where $\mathbf{D_h} \in \mathbb{R}^{m \times n}$ is the Jacobian of observation function $h(\cdot)$. 

The Kalman gain matrix decides the influence of measurements on the estimated state. When the measurement error covariance $\mathbf{R}_{k+1}$ approaches zero, the Kalman gain $\mathbf{K}$ gives more weight to the residual defined as $(\mathbf{z}_{k+1}-h(\mathbf{x}^f_{k+1})$. On the other hand, when the error covariance $\mathbf{P}^f_{k+1}$ is very small, the Kalman gain $\mathbf{K}$ weights the residual less heavily. Each row of the Kalman gain matrix contains the influence of all observation points on one element of the state $\mathbf{x}_{k+1}$ corresponding to that row. The analyzed error covariance matrix is calculated using Equation~\ref{eq:ekf_pa}, where $\mathbf{I} \in \mathbb{R}^{n \times n}$ is an identity matrix.

\subsection{Ensemble Kalman filter}
When the system is high-dimensional, i.e., $n$ is very large, then the computations for the EKF algorithm are practically infeasible. In addition, the EKF algorithm requires computation of Jacobians and it might be numerically difficult to compute Jacobians for complex models. Ensemble filtering techniques are attractive for such systems where the approximate state of the system is estimated using the standard Monte Carlo framework. 

In EKF, the mean estimate of the state $\widehat{\mathbf{x}}_k$ and the error covariance matrix $\widehat{\mathbf{P}}_{k}$ are updated sequentially. In contrast to an EKF algorithm, we apply the forecast step to an ensemble of states in the EnKF algorithm\cite{lewis2006dynamic}. The sample mean and covariance of the ensembles analyses represent the analyzed state estimate $\widehat{\mathbf{x}}_k$ and error covariance matrix $\widehat{\mathbf{P}}_{k}$.  

Let $x_0$ be an initial condition drawn from the Gaussian distribution with mean $\mathbf{m}_0$ and the covariance matrix $\mathbf{P}_0$, i.e., $x_0 \sim {\cal{N}}(\mathbf{m}_0,\mathbf{P}_0)$. In our notation we use $\mathbf{X}_k(i)$ to denote the $i^{\text{th}}$ member of ensembles and $N$ is the size of ensembles, i.e., $i=1,2,\dots,N$. The procedure for the EnKF is summarized in Algorithm~\ref{alg:enkf}. We initialize the state of the system for all ensemble members from known distribution of the initial condition for the system as given in Equation~\ref{eq:enkf_xo}. Then, we forecast the state of the system for all ensemble members between two observation points (i.e., from time $t_k$ to $t_{k+1}$) using the nonlinear model dynamics as given in Equation~\ref{eq:enkf_xk}. The forecast state estimate and error covariance are calculated based on the sample mean and sample variance of all ensembles as given in Equation~\ref{eq:enkf_xf} and Equation~\ref{eq:enkf_pf}, respectively. 

\begin{figure}
\begin{algorithm}[H]
\caption{Ensemble Kalman filter}
\label{alg:enkf}
\begin{algorithmic}[1]
\STATE Initialize the state of the system for different ensemble members. \\
\begin{align}
    \mathbf{\widehat{X}}_0(i) &= \mathbf{m}_0 + \mathbf{y}_0(i), \label{eq:enkf_xo}\\
\end{align}
where $\mathbf{y}_0(i) \sim N(0,\mathbf{P}_0)$.
\STATE For $k=0,1,\dots$ proceed as follow
\begin{itemize}
    \item Forecast step: 
    \begin{itemize}
        \item Integrate the state estimate all ensemble members from time $t_k$ to $t_{k+1}$ as follow 
        \begin{align}
        \mathbf{X}^f_{k+1}(i) &= \mathbf{M}(\widehat{\mathbf{X}}_k(i)) + \mathbf{w}_{k+1}.
        \label{eq:enkf_xk}
        \end{align}
        \item Compute the sample mean and error covariance as follow
        \begin{align}
            \mathbf{x}^f_{k+1} &= \frac{1}{N} \sum_{i=1}^N \mathbf{X}^f_{k+1}(i), \label{eq:enkf_xf}\\
        \mathbf{E}^f_{k+1}(i) &= \mathbf{X}^f_{k+1}(i) - \mathbf{x}^f_{k+1}, \\
        \mathbf{P}^f_{k+1} &= \frac{1}{N-1} \sum_{i=1}^N  \mathbf{E}^f_{k+1}(i) [\mathbf{E}^f_{k+1}(i)]^{\text{T}}. \label{eq:enkf_pf}
        \end{align}
    \end{itemize}
    \item Data assimilation step: 
    \begin{itemize}
        \item Once the observations are available at time $t_{k+1}$, generate $N$ realizations of virtual observations as follow
        \begin{align}
            {\mathbf{Z}}_{k+1}(i) &= {\mathbf{z}}_{k+1} + \mathbf{v}_{k+1}(i), \label{eq:enkf_zv}
        \end{align} 
        where $\mathbf{v}_{k+1}(i) \sim N(0,\mathbf{R}_{k+1})$.
        \item Assimilate the state estimate with virtual observations for all ensemble members as follow
        \begin{align}
            \widehat{\mathbf{X}}_{k+1}(i) &= \mathbf{X}^f_{k+1}(i) + \mathbf{K}[\mathbf{Z}_{k+1}(i) - h(\mathbf{X}^f_{k+1}(i))], \label{eq:enkf_xe}\\
            \mathbf{K} &= \mathbf{P}^f_{k+1} \mathbf{D}^\text{T}_\mathbf{h}[ \mathbf{D}_\mathbf{h} \mathbf{P}^f_{k+1} \mathbf{D}^\text{T}_\mathbf{h} + \mathbf{R}_{k+1}]^{-1}.
        \end{align}
        \item Compute the sample mean to get analysis state estimate at time $t_{k+1}$
        \begin{align}
            \widehat{\mathbf{x}}_{k+1} &= \frac{1}{N} \sum_{i=1}^N \widehat{\mathbf{X}}_{k+1}(i). \label{eq:enkf_xa}
        \end{align}
    \end{itemize} 
\end{itemize}
\end{algorithmic}
\end{algorithm} 
\end{figure}

Once the observations $\mathbf{z}_{k+1}$ are available at time $t_{k+1}$, we create $N$ different virtual observations using Equation~\ref{eq:enkf_zv}. In the original formulation of EnKF algorithm proposed by Evensen\cite{evensen1994sequential}, virtual observations were not used in the assimilation step. However, Burgers et al.\cite{burgers1998analysis} showed that it is essential to include random perturbations to observations to ensure that the analyzed covariance is not underestimated. Once the virtual observations are generated, the forecast state estimate for all ensembles are assimilated using Equation~\ref{eq:enkf_xe}. The Kalman gain $\mathbf{K}$ is computed using the same formula as the EKF algorithm. The analysis state estimate is calculated using the sample mean of analyzed state estimate for all ensemble members as given in Equation~\ref{eq:enkf_xa}. 

\subsection{Deterministic ensemble Kalman filter}
Sakov et al.\cite{sakov2008deterministic} proposed a modification in traditional EnKF that results into matching the analyzed error covariance to that of standard Kalman filter without the need to virtual observations.

\begin{figure}
\begin{algorithm}[H]
\caption{Deterministic ensemble Kalman filter}
\label{alg:denkf}
\begin{algorithmic}[1]
\STATE Initialize the state of the system for different ensemble members. \\
\begin{align}
    \mathbf{\widehat{X}}_0(i) &= \mathbf{m}_0 + \mathbf{y}_0(i), \label{eq:denkf_x0}\\
\end{align}
where $\mathbf{y}_0(i) \sim N(0,\mathbf{P}_0)$.
\STATE For $k=0,1,\dots$ proceed as follow
\begin{itemize}
    \item Forecast step: 
    \begin{itemize}
        \item Integrate the state estimate all ensemble members from time $t_k$ to $t_{k+1}$ as follow 
        \begin{align}
        \mathbf{X}^f_{k+1}(i) &= \mathbf{M}(\widehat{\mathbf{X}}_k(i)) 
        \end{align}
        \item Compute the sample mean, ensemble anomalies, and error covaraince as follow
        \begin{align}
            \mathbf{x}^f_{k+1} &= \frac{1}{N} \sum_{i=1}^N \mathbf{X}^f_{k+1}(i), \\
        \mathbf{A}^f_{k+1}(i) &= \mathbf{X}^f_{k+1}(i) - \mathbf{x}^f_{k+1}, \label{eq:denkf_an}\\
        \mathbf{P}^f_{k+1} &= \frac{1}{N-1} \sum_{i=1}^N  \mathbf{A}^f_{k+1}(i) [\mathbf{A}^f_{k+1}(i)]^{\text{T}} \label{eq:denkf_Pf}.
        \end{align}
    \end{itemize}
    \item Data assimilation step: 
    \begin{itemize}
        \item Once the observations are available at time $t_{k+1}$, assimilate the forecast state estimate with the observation as follow
        \begin{align}
            \widehat{\mathbf{x}}_{k+1} &= \mathbf{x}^f_{k+1} + \mathbf{K}[\mathbf{z}_{k+1} - h(\mathbf{x}^f_{k+1})], \label{eq:denkf_xa}\\
            \mathbf{K} &= \mathbf{P}^f_{k+1} \mathbf{D}^\text{T}_\mathbf{h}[ \mathbf{D}_\mathbf{h} \mathbf{P}^f_{k+1} \mathbf{D}^\text{T}_\mathbf{h} + \mathbf{R}_{k+1}]^{-1} \label{eq:denkf_KG}.
        \end{align}
        \item Compute the analyzed anomalies as below
        \begin{align}
            \widehat{\mathbf{A}}_{k+1}(i) = \mathbf{A}^f_{k+1}(i) - \frac{1}{2}\mathbf{K} \mathbf{D}_\mathbf{h}\mathbf{A}^f_{k+1}(i). \label{eq:denkf_aa}
        \end{align}
        \item Calculate the analyzed ensemble using the analyzed state estimate and analyzed anomalies as follow
        \begin{align}
            \widehat{\mathbf{X}}_{k+1}(i) = \widehat{\mathbf{A}}_{k+1}(i) + \widehat{\mathbf{x}}_{k+1}. \label{eq:denkf_ea}
        \end{align}
    \end{itemize} 
\end{itemize}
\end{algorithmic}
\end{algorithm}
\end{figure}

The procedure for the deterministic EnKF (DEnKF) is summarized in Algorithm~\ref{alg:denkf}. In practice (e.g., when $n>>N$), we compute Equation~\ref{eq:denkf_KG} using its square root version (without storing or computing $\mathbf{P}^f_{k+1}$ explicitly) as follows 
\begin{align}
   \mathbf{K} = \frac{1}{N-1}\mathbf{A}^f(\mathbf{D}_\mathbf{h}\mathbf{A}^f)^{\text{T}}\left[\frac{1}{N-1}(\mathbf{D}_\mathbf{h}\mathbf{A}^f)(\mathbf{D}_\mathbf{h}\mathbf{A}^f)^\text{T} + \mathbf{R_{k+1}}\right]^{-1}
\end{align}
where a size of $\mathbb{R}^{n \times N}$ matrix is concatenated as follows
\begin{align}
\mathbf{A}^f = [\mathbf{A}^f_{k+1}(1), \mathbf{A}^f_{k+1}(2), \dots, \mathbf{A}^f_{k+1}(N)].
\end{align}
In other words, we skip computing Equation~\ref{eq:denkf_Pf}, and use its reduced-rank square root definition given by
\begin{align}
\mathbf{P}^f_{k+1} = \frac{1}{N-1}\mathbf{A}^f (\mathbf{A}^f)^T.
\end{align}

We start the DEnKF algorithm in a similar manner as the EnKF algorithm by initializing the state estimate for all ensemble members using Equation~\ref{eq:denkf_x0}. The anomalies between the forecast estimate of all ensembles and its sample mean is computed utilizing Equation~\ref{eq:denkf_an}. Once the observations are available at time $t_{k+1}$, the forecast state estimate is assimilated as given in Equation~\ref{eq:denkf_xa}, where the Kalman gain $\mathbf{K}$ is computed in a similar manner as the EKF algorithm. The anomalies for all ensemble members are updated separately with half the Kalman gain. Therefore, the analyzed anomalies for all ensemble members are calculated using Equation~\ref{eq:denkf_aa}. The analyzed state estimate for all ensembles members are obtained by offsetting the analyzed anomalies with the analyzed state estimate and is computed using Equation~\ref{eq:denkf_ea}.   

\section{Nudging dynamics} \label{sec:nudging}

Nudging is another data assimilation method that was introduced by Anthes\cite{anthes1974data} for initialization of hurricane models from real observational data. Contrary to variational and sequential data assimilation methods that minimize the cost function based on the error between model forecast and observations, nudging methods utilize the forecast error as a constraint to the model evolution equation. The evolution of the dynamical system based on nudging methods can be written as 
\begin{equation}\label{eq:nudging}
    \mathbf{x}_{k+1} = \textbf{M}(\mathbf{x}_{k}) + G_k \mathbf{e}_{k},
\end{equation}
where $G_k \in \mathbb{R}^{n \times m}$ is called the time varying nudging coefficient matrix. The forecast error $\mathbf{e}_{k}$ in Equation~\ref{eq:nudging} is computed as below
\begin{equation}
    \mathbf{e}_{k} = \mathbf{z}_k - h(\mathbf{x}_k).
\end{equation}
The correction term in Equation~\ref{eq:nudging} is proportional to $\mathbf{e}_{k} \in \mathbb{R}^{m}$ (i.e., in the observation space) and therefore this form of nudging is called as observation nudging. The literature on nudging can be divided into different classes/versions based on how the nudging coefficient matrix is computed. Lakshmivarahan et al.\cite{lakshmivarahan2013nudging} offer an overview of theoretical aspect of nudging methods and present promising directions of research on the nudging process of dynamic data assimilation. Nudging methods have been applied for different applications such as forecast of Indian Monsoon \cite{krishnamurti1991physical}, diagnostic studies of mesoscale processes in mid-latitude weather systems \cite{stauffer1990use,stauffer1991use}, and operational predictions in meteorology and oceanography \cite{lorenc1991meteorological, derber1989global}.   
Zou et al.\cite{zou1992optimal} proposed a parameter-estimation approach to obtain optimal nudging coefficients using a variational data assimilation method. They estimated the parameters of the nudging coefficient matrix by solving the constrained minimization problem utilizing the Lagrangian formulation. Their cost function consists of two parts, the first part corresponds to the misfit between the model results and observations, and the second part was related to keeping the new estimate of nudging coefficients close to its prior estimate. They enforced the nudged dynamics given in Equation~\ref{eq:nudging} as a strong constraint to the optimization problem. They demonstrated the performance of optimal nudging method for an adiabatic version of the National Meteorological Center (NMC) spectral model with eighteen vertical layers. Vidard et al.\cite{vidard2003determination} introduced another approach to estimate optimal nudging coefficient matrix using the Kalman filter. They illustrated the proposed approach for Burgers equation and shallow-water equations in a twin experiment framework and showed noticeable improvement in the prediction.                
Auroux et al.\cite{auroux2005back} introduced the back and forth nudging (BFN) algorithm where the set of observations are incorporated into the model by running it forward in time, starting with some initial condition. After the forward run is completed, the model is again run backward in time, starting from the final state obtained by the standard nudging method. During the backward integration the use of opposite sign for the nudging term as to forward integration makes this algorithm numerically stable. This procedure is repeated in BFN algorithm until the convergence. Therefore, it helps to reduce forecast error on a finite time window. One of the advantage of the BFN algorithm is that it does not require the linearization of nonlinear equations in order to have the adjoint model or to solve any optimization problem. The BFN algorithm was tested for the Lorenz-63 model and for quasi-geostrophic model in the presence of perfect and noisy observations \cite{auroux2008nudging} and showed comparable prediction to the 4D-VAR algorithm.

Spectral nudging is another technique where the nudging term is added in the spectral domain with maximum efficiency for large scales and no effect for small scales \cite{waldron1996sensitivity}. This method has been successfully applied to force large-scale atmospheric states from global climate models onto a regional climate model \cite{von2000spectral,radu2008spectral,miguez2004spectral,rockel2008dynamical,schubert2017optimal}. The main idea in spectral nudging is that small-scale details for weather prediction are governed by the interplay between larger-scale atmospheric flow and geographic features like mountains, and land-sea distribution. It is computationally impractical to resolve these small scales in global climate models. Therefore, spectral nudging is applied to match overlapping scales in global and regional climate models by forcing the regional model to behave as global model. Spectral nudging method has also been applied for inferring flow parameters for turbulent flows \cite{di2018inferring}, and for three-dimensional homogeneous isotropic turbulence \cite{di2020synchronization}. There are also nudging methods that make use of present and past observations in the formulation of forcing term to drive the model evolution toward observation\cite{rey2014accurate,pazo2016data}. An et al.\cite{an2017estimating} used the time delayed nudging method \cite{rey2014accurate} for estimating the state of geophysical system from sparse observation data.  

\section{Long short-term memory nudging} \label{sec:LSTM}

With the huge amount of data generated from high-fidelity numerical simulations, non-invasive experimental techniques like particle image velocimetry (PIV), and satellite data, there is a growing interest in using machine learning for data assimilation \cite{abarbanel2018machine,gilbert2010machine}. One of the difficulties in weather and climate prediction is that atmospheric flows are multiscale in nature and their dynamics are typically chaotic. Several data-driven algorithms can address these challenges. The recurrent neural networks (RNNs) are particularly attractive for complex dynamical systems due to their ability to capture temporal dependencies and to take state history into account for future state prediction. One of the problems with RNN is that the gradient vanishes during the learning procedure. Long short-term memory \cite{hochreiter1997long} is a type of RNN that alleviates this issue of vanishing gradient \cite{hochreiter1998vanishing} by employing cell architecture that remembers or forgets information. 

There is a rich literature on the application of LSTM for modeling chaotic dynamical systems. Vlachas et al.\cite{vlachas2018data} proposed a data-driven forecasting method for the high-dimensional chaotic system by modeling their temporal dynamics on reduced order space using LSTM. They also integrated the LSTM with a mean stochastic model to ensure convergence and demonstrated its improved prediction performance compared to the Gaussian process. In Wan et al.\cite{wan2018data}, the LSTM was employed to learn the mismatch between imperfect Galerkin based reduced order model and the actual dynamics projected onto the reduced order space. They showed the improved performance of the proposed framework for the prediction of extreme events. Jia et al.\cite{jia2020physics} introduced the physics-guided RNN that combines the LSTM and physics-based model to model the dynamics of temperature in lakes. They utilized a physics-based regularization as a penalty term to the optimization cost function to enforce physics into the training. Apart from LSTM, other machine learning algorithms such as reservoir computing have been used for modeling chaotic dynamical systems \cite{pathak2017using,pathak2018model} and residual network for predicting dynamical system evolution \cite{chashchin2019predicting,chen2020generalized}. In a recent study, Vlachas et al.\cite{vlachas2020backpropagation} investigated the performance of LSTM trained with backpropagation through time and reservoir computing for long term forecasting of chaotic dynamical systems.       

Zhang et al.\cite{zhang2019lstm} presented an LSTM based Kalman filter for data assimilation of two-dimensional spatio-temporal varying depth of ocean field for underwater glider path planning. In their study, the temporal evolution of spatial basis function was modeled using LSTM. They train the LSTM network to predict the future temporal coefficients based on the historical states of these coefficients. Jin et al.\cite{jin2019machine} utilized LSTM to perform observation bias correction for data assimilation of dust storm prediction. They showed that with the LSTM model for bias corrections, existing measurements are used precisely and that improves the resulting prediction accuracy. In the work by Loh et al.\cite{loh2018deep}, the LSTM was deployed as a prediction model for their EnKF approach to achieve real-time production forecast in natural gas wells. Xingjian et al.\cite{xingjian2015convolutional} proposed a convolutional LSTM framework to predict the rainfall intensity over a short period of time and illustrated its ability to capture more improved correlation than existing methods. 

Motivated by the previous successes of employing neural networks for making better predictions in geophysical applications, in the present study, we introduce an LSTM nudging scheme. The LSTM network is trained to learn the correction term based on the background state of the system and observations. To train the LSTM network, we initialize the state of the system for different training sets from prior distribution of the true initial state. This step is similar to initializing different ensemble members in the case of the EnKF algorithm. We then evolve the system with erroneous initial conditions and compute the correction term using Equation~\ref{eq:lstm_corr} at all observation points. The input features to the LSTM network (denoted by $\mathcal{X}_k$) consists of full state of the system (from erroneous initial conditions) and current observations, i.e., $\mathcal{X}_k = \{\mathbf{\widehat{X}}_k(i);\mathbf{z}_k \} \in \mathbb{R}^{n+m}$, where $m$ is number of observations. Based on these input features, the LSTM is trained to learn the correction term for all states, i.e., the output of the LSTM is $\mathcal{Y}_k = \boldsymbol{\epsilon}_k(i) \} \in \mathbb{R}^{n}$. The LSTM network is capable of capturing the temporal dependencies and utilize it to forecast the system's future state. Therefore, we can also train the LSTM network by including the temporal history of the system's states and observations as input features. Readers are referred to Rahman et al.\cite{rahman2019non} for further details on incorporating the temporal history of the system's state into training. The procedure for training phase of the LSTM nudging scheme is outlined in Algorithm~\ref{alg:lstmt} 

\begin{figure}
\begin{algorithm}[H]
\caption{LSTM Nudging (Training phase)}
\label{alg:lstmt}
\begin{algorithmic}[1]
\STATE Initialize the state of the system for different training sets from prior distribution of $\mathbf{x}_0 \sim N(\mathbf{m}_0,\mathbf{P}_0)$. \\
\begin{align}
    \mathbf{\widehat{X}}_0(i) &= \mathbf{m}_0 + \mathbf{y}_0(i), \label{eq:lstmt_x0}
\end{align}
where $\mathbf{y}_0(i) \sim N(0,\mathbf{P}_0)$.
\STATE Integrate the dynamical system and store the system's state at all observation points, i.e., at time $t_1, \dots, t_k$
\STATE Compute the correction term at time $t_1, \dots, t_k$ with respect to the true state of the system as follow
\begin{align}\label{eq:lstm_corr}
    \boldsymbol{\epsilon}_k(i) = \mathbf{\widehat{X}}_k(i) - \mathbf{\overbar{x}}_k,
\end{align}
where $\mathbf{\overbar{x}}_k$ is the true state of the system at $t_k$.
\STATE Each sample of the input training matrix $\mathcal{X}_k$ and corresponding output data matrix $\mathcal{Y}_k$ is constructed as follow
\begin{align}
    \mathcal{X}_k &= \{\mathbf{\widehat{X}}_k(i);\mathbf{z}_k \} \in \mathbb{R}^{n+m}, \\
    \mathcal{Y}_k &= \{\boldsymbol{\epsilon}_k(i) \} \in \mathbb{R}^{n},
\end{align}
where $m$ is the number of observations.
\STATE Train the LSTM model to learn the mapping from input to output
\begin{align}
    \mathcal{M}: \mathcal{X}_k \Rightarrow \mathcal{Y}_k.
\end{align}
\end{algorithmic}
\end{algorithm}
\end{figure}

We adopt the predictor-corrector approach during online deployment. Since the LSTM network is trained to learn the mapping from the state of the system generated with the erroneous initial condition, we start with two systems. We use the superscript $E$ to denote the system with erroneous initial condition and $C$ to denote the evolution of the system whose state is corrected at each observation point. The procedure for online deployment is reported in Algorithm~\ref{alg:lstmd}. We start with initializing two systems with the same initial condition based on some educated guess. The dynamics of erroneous and corrected systems is evolved simultaneously as given in Equation~\ref{eq:lstmd_xke} and Equation~\ref{eq:lstmd_xkc}, respectively. Once the observations are available, we determine the correction term using the trained LSTM network as shown in Equation~\ref{eq:lstmd_c}. This correction is for the state of the system generated with the erroneous initial condition. Therefore, the correction is added to the erroneous system's state at that time and assigned to the corrected system. Between two observation points, the corrected system is evolved using this nudged state estimate and we will show that it follows the same trajectory as the true system in Section~\ref{sec:results}. 

\begin{figure}
\begin{algorithm}[H]
\caption{LSTM Nudging (Online deployment)}
\label{alg:lstmd}
\begin{algorithmic}[1]
\STATE Initialize the state of the system for two members with an educated guess for an initial condition. \\
\begin{align}\label{eq:ldtmd_x0}
    \mathbf{{X}}^E_0 &= \mathbf{x}_0, \\
    \mathbf{{X}}^C_0 &= \mathbf{x}_0.
\end{align}
\STATE For $k=0,1,\dots$ proceed as follow
\begin{itemize}
    \item Forecast step: Integrate the state estimate for two systems from time $t_k$ to $t_{k+1}$ as follow 
    \begin{align}
        \mathbf{X}^E_{k+1} &= \mathbf{M}({\mathbf{X}}^E_k), \label{eq:lstmd_xke}\\
        \mathbf{X}^C_{k+1} &= \mathbf{M}({\mathbf{X}}^C_k). \label{eq:lstmd_xkc}
    \end{align}
    \item Data assimilation step: Once the observations are available at time $t_{k+1}$, they are used to determine the correction term with the trained LSTM network and correct the state estimate as follow
    \begin{align}
        \boldsymbol{\epsilon}_{k+1} = \mathcal{M}(\{\mathbf{X}^E_{k+1};\mathbf{z}_{k+1}\}), \label{eq:lstmd_c}\\
        \mathbf{X}^C_{k+1} = \mathbf{X}^E_{k+1} + \boldsymbol{\epsilon}_{k+1}.
    \end{align}
\end{itemize}
\end{algorithmic}
\end{algorithm}
\end{figure}

We highlight some of the features of the LSTM nudging framework here. The LSTM nudging framework is highly modular and it can be implemented with other types of neural network architectures also based on the size or type of problems. For example, convolutional autoencoders are gaining popularity to find the nonlinear basis functions of complex physical systems and they are complemented with the LSTM network for learning the latent-space dynamics \cite{gonzalez2018deep,mohan2019compressed,maulik2020reduced,lee2020model,erichson2019physics,agostiniexploration,wu2020data}. The LSTM nudging framework can be easily applied to high dimensional systems, where convolutional autoencoders are employed for dimensionality reduction and the LSTM is trained to learn the nudging dynamics in latent-space instead of high-dimensional space. Novel neural network architectures like generative adversarial networks (GANs)\cite{goodfellow2014generative,amirian2019data} can also be applied to learn the nudging dynamics. Another feature of the LSTM nudging scheme is that once the network is trained with the archival or background data, it can be retrained efficiently with transfer learning as the new observation data becomes available. Therefore, training the LSTM network for the first time is the only computationally heavier part of the LSTM nudging scheme. Our main goal in this study is to illustrate that the neural network can be effectively trained to provide accurate and stable nudging dynamics.


\begin{figure*}
\centering
\mbox{\subfigure{\includegraphics[width=0.85\textwidth]{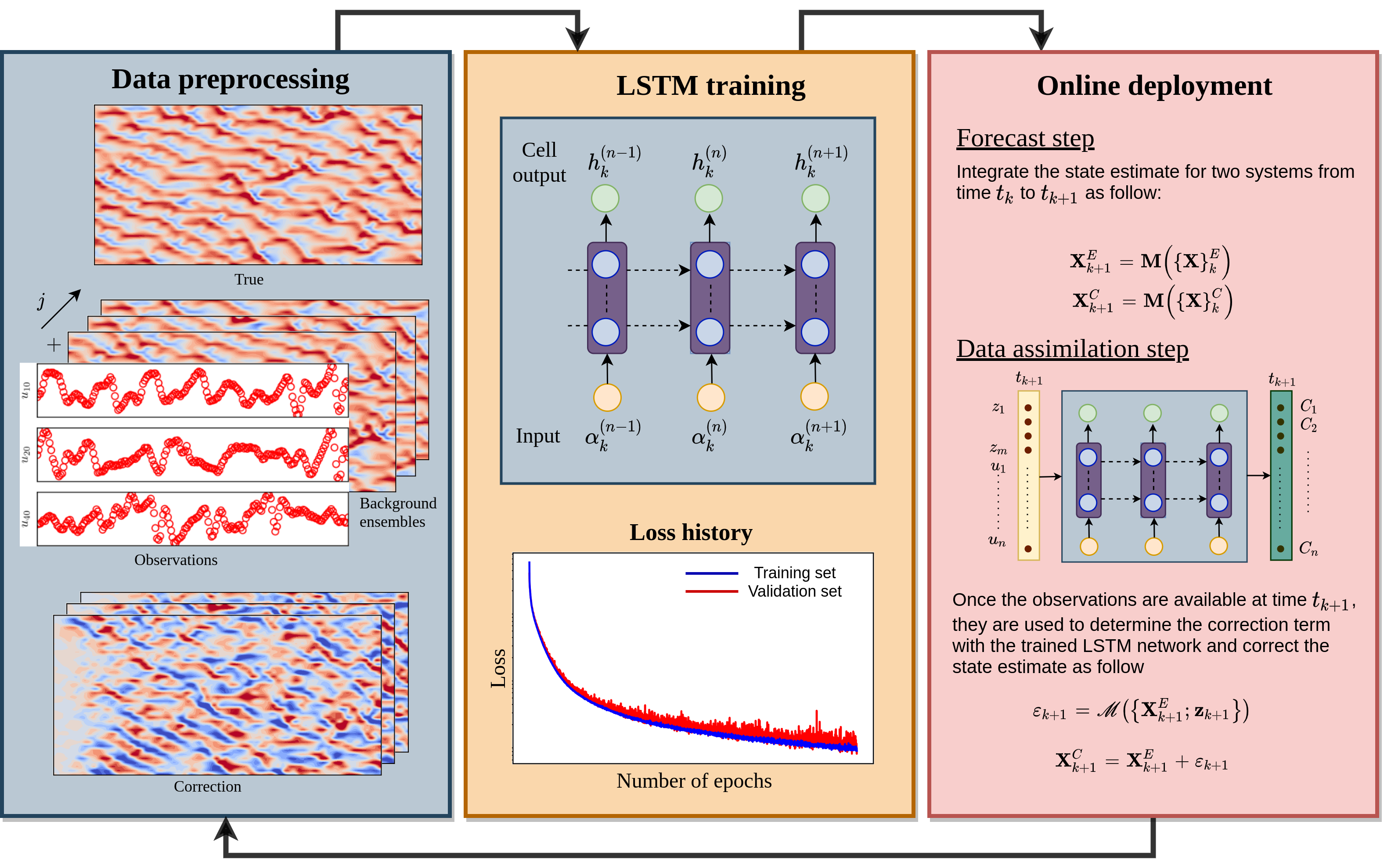}}}
\caption{Overview of the LSTM-DA framework. The LSTM-DA framework consists of three main steps: data preprocessing, training the neural network, and the deployment of trained network.} 
\label{fig:da_lstm}
\end{figure*}

\section{Data assimilation problem set-up} \label{sec:L96}

In this section we describe the Lorenz 96 model proposed by Lorenz \cite{lorenz1996predictability}, which is commonly used as a prototypical test case in data assimilation. This model describes the temporal evolution of atmospheric quantity discretized spatially over a single latitude circle. The system of ordinary differential equation governing the Lorenz 96 model can be written as 
\begin{equation}
    \frac{d u_i}{dt} = u_{i-1} (u_{i+1} - u_{i-2}) - u_i + F,
    \label{eq:l96}
\end{equation}
for $i \in \{1,2,\dots,n\}$. The first term on the right hand side of Equation~\ref{eq:l96} is the nonlinear advection term, the second term present an internal dissipation, and the third term present an external forcing. We use $n=40$ and $F=10$ in our analysis. We apply the periodic boundary conditions at ghost points, i.e., $u_0=u_n, u_{-1}=u_{n-1},$ and $u_{n+1}=u_1$.

We use the fourth-order Runge-Kutta scheme for time integration with a time step of $\Delta t = 0.005$. To generate a physical initial condition for the forward run, we start with an equilibrium condition at time $t=-5$. The equilibrium condition for the model is $u_i=F$ for $i \in \{1,2,\dots,n\}$. We introduce a very little perturbation to the equilibrium state for the state $u_{20}$, i.e., we set $u_{20}=F+0.01$ to generate chaotic dynamics and then do the time integration up to $t=0$. Once the true initial condition is generated, we run the forward solver up to time $t=10$.

The twin experiment is one of the most commonly used methods to validate any data assimilation algorithm before it can be applied to real-life applications \cite{abarbanel2013predicting}. For twin experiments, first, we generate the $n$-dimensional data for the Lorenz 96 model and select $m$ observations. These observations are obtained by adding some noise to the true state of the system to take experimental uncertainties and measurement error into account. The observations are also sparse in time, meaning that the time interval between two observations can be different from the time step of the model. For our twin experiments, we assume that observations are recorded at every $10^{\text{th}}$ time step of the model. Therefore, the time difference between two observations is $\delta t = 0.05$. The analysis time step $\delta t = 0.05$ is representative of six hours of a data assimilation cycle of global meteorological models. The accurate estimation of the full state of the system depends upon the number of observations that are assimilated by the model \cite{whartenby2013number}. We assume that observations locations are constant throughout the time unlike asynchronous observations where they can be rotated \cite{fertig2007comparative}. We compare the performance of traditional data assimilation algorithms and the proposed LSTM nudging algorithm for three sets of observations. The first set of observations is very sparse with only 10\% of the full state of the system (i.e., $m=4$), utilizing observations for states $[u_{10},u_{20},u_{30},u_{40}] \in R^4$. In a second set of observations ($m=8$), we employ observations at $[u_5,u_{10},\dots,u_{40}] \in R^8$ for the assimilation. The third set of observations consists of 50\% of the full state of the system ($m=20$), i.e., observations at states $[u_2,u_{4},\dots,u_{40}] \in R^{20}$ for the assimilation. 


\section{Results}
\label{sec:results}
In this section, we describe the results of numerical experiments with the Lorenz 96 model using algorithms discussed in Section~\ref{sec:kalman},~\ref{sec:nudging} and~\ref{sec:LSTM}. 
We assume that our model is perfect for all numerical experiments except for the EKF and EnKF algorithm. For these two algorithms, it is found that an introduction of small uncertainty in the model provides more accurate predictions than the assumption of a perfect model. For the aforementioned two algorithms, we assume that the model noise is drawn from the Gaussian distribution with zero mean and variance $1\times 10^{-4}$. The observations are created by adding random noise from Gaussian distribution with zero mean and variance $1 \times 10^{-2}$ to the true state of the system. The erroneous initial condition is generated by adding a noise form Gaussian distribution of zero mean and $1 \times 10^{-2}$ variance to the true initial condition. To ensure a fair comparison between EnKF and DEnKF, we use an equal number of ensembles in both algorithms. For the comparison, we plot time evolution of states $u_{10},u_{21},u_{39}$, and also the full state trajectory of the Lorenz 96 model. We use black lines to denote true states, dashed blue lines to denote states with the erroneous initial condition and dashed-dotted green color lines for assimilated states. The observations for the state $u_{10}$ are shown with red circles in all of the time series plots. 

\begin{figure*}
\centering
\mbox{\subfigure{\includegraphics[width=0.96\textwidth]{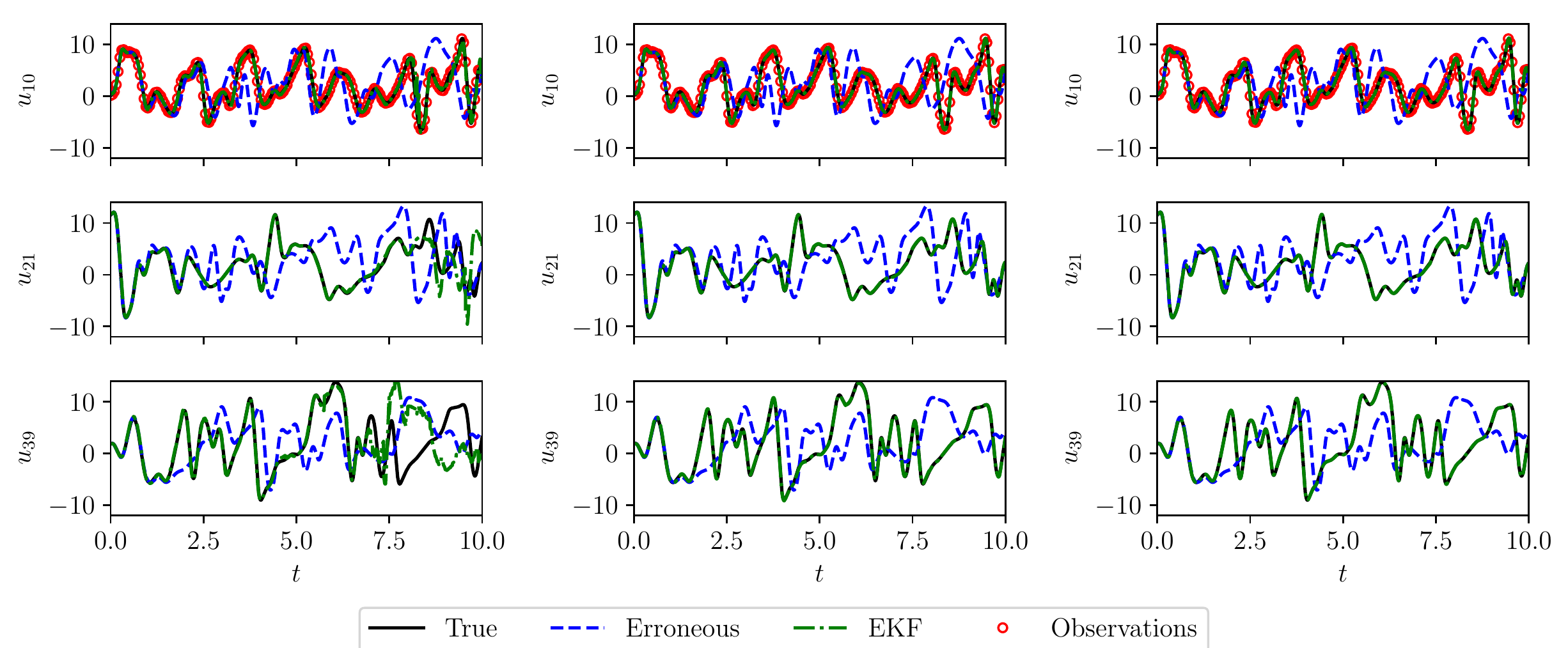}}}
\caption{Selected trajectories of the Lorenz 96 model with the analysis performed by the extended Kalman filter (EKF) using observations from $m=4$ (left), $m=8$ (middle), and  $m=20$ (right) state variables at every 10 time steps.} 
\label{fig:ekf_3}
\end{figure*}

\begin{figure*}
\centering
\mbox{\subfigure{\includegraphics[width=0.96\textwidth]{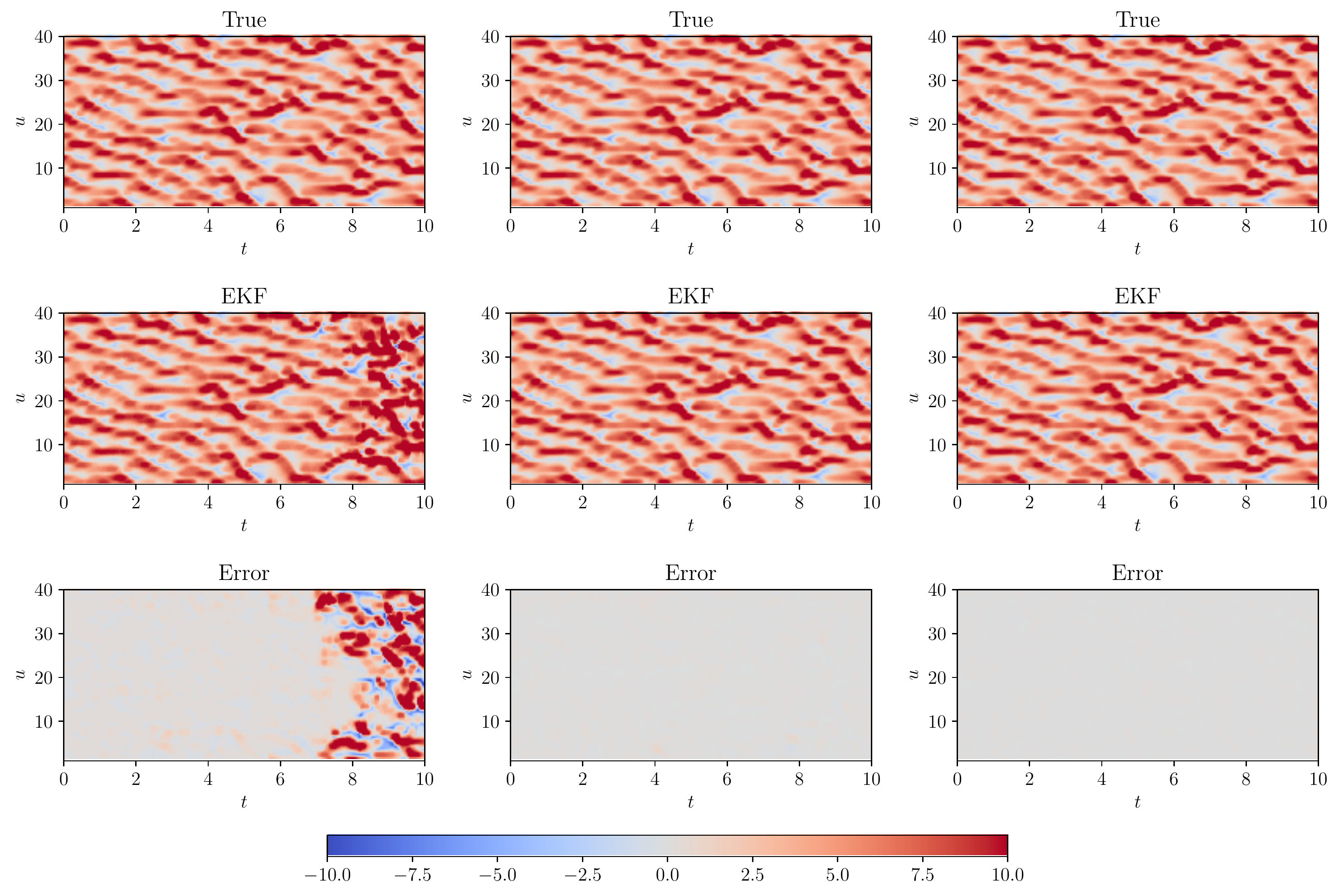}}}
\caption{Full state trajectory of the Lorenz 96 model with the analysis performed by the extended Kalman filter (EKF) using observations from $m=4$ (left), $m=8$ (middle), and  $m=20$ (right) state variables at every 10 time steps.} 
\label{fig:ekf_all}
\end{figure*}

\begin{figure*}
\centering
\mbox{\subfigure{\includegraphics[width=0.96\textwidth]{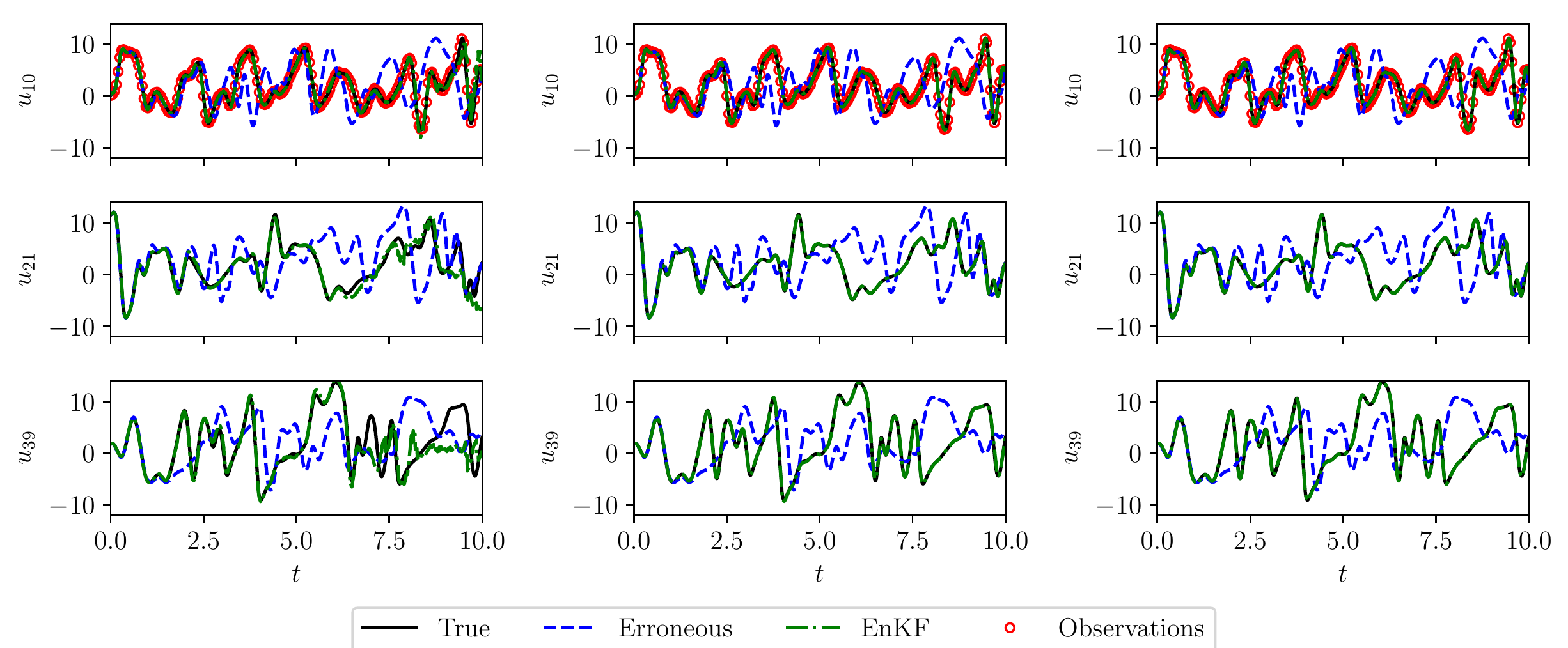}}}
\caption{Selected trajectories of the Lorenz 96 model with the analysis performed by the ensemble Kalman filter (EnKF) with $N=40$ member ensemble using observations from $m=4$ (left), $m=8$ (middle), and  $m=20$ (right) state variables at every 10 time steps.} 
\label{fig:enkf_3}
\end{figure*}

\begin{figure*}
\centering
\mbox{\subfigure{\includegraphics[width=0.96\textwidth]{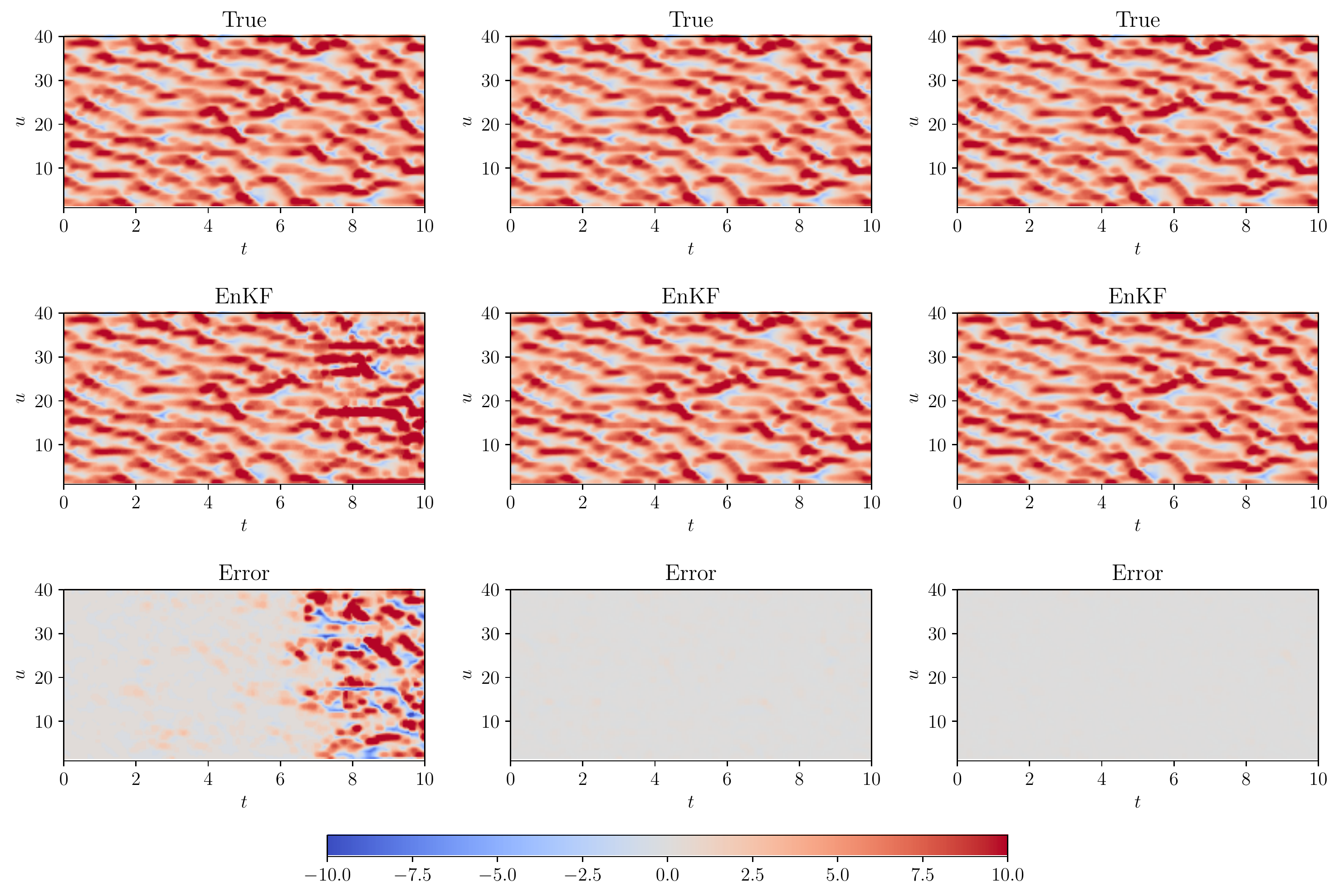}}}
\caption{Full state trajectory of the Lorenz 96 model with the analysis performed by the ensemble Kalman filter (EnKF) with $N=40$ member ensemble using observations from $m=4$ (left), $m=8$ (middle), and  $m=20$ (right) state variables at every 10 time steps.} 
\label{fig:enkf_all}
\end{figure*}

In Figure~\ref{fig:ekf_3}, we present the time evolution of selected states for three different number of observations included in the assimilation of the EKF algorithm. There is an excellent agreement between true and assimilated states $u_{21}$ and $u_{39}$ when more than 20\% observations are utilized for assimilation. We also provide the full state trajectory of the Lorenz 96 model in Figure~\ref{fig:ekf_all}. The results obtained clearly show that the EKF algorithm can determine the correct state trajectory with more than 20\% observations, i.e., for $m \ge 8$. We observe a discrepancy in prediction after $t \sim 7$, when only four observations are used in the assimilation step. Figure~\ref{fig:enkf_3} shows the time evolution of selected states predicted by the EnKF algorithm with $N=40$ ensemble members. We notice some discrepancy between the true and predicted states with $m=12$ observations after $t \sim 7.5$. If we compare the full state trajectory prediction by the EnKF algorithm in Figure~\ref{fig:enkf_all}, we can conclude that there is almost a perfect match between true and assimilated states with more than 8 observations. Since the EnKF algorithm is based on the Monte Carlo framework, its accuracy can be improved by applying increased number of ensembles. The typical number of ensembles is $O(100)$ for high-dimensional systems \cite{houtekamer2001sequential,ott2004local,jardak2010comparison}. Considering that the Lorenz 96 model is a lower-dimensional system with $n=40$ states, we apply only 40 ensemble members. If we consider the computational cost of the EKF algorithm, the major bottleneck is the propagation of the error covariance matrix as given in Equation~\ref{eq:ekf_pk}. The computational overhead of the EnKF algorithm goes up with an increase in the number of ensembles. However, with the advancement in parallel algorithms and high-performance computing, ensemble Kalman filter algorithms are particularly attractive data assimilation of complex physical systems \cite{anderson2007scalable}. 

\begin{figure*}
\centering
\mbox{\subfigure{\includegraphics[width=0.96\textwidth]{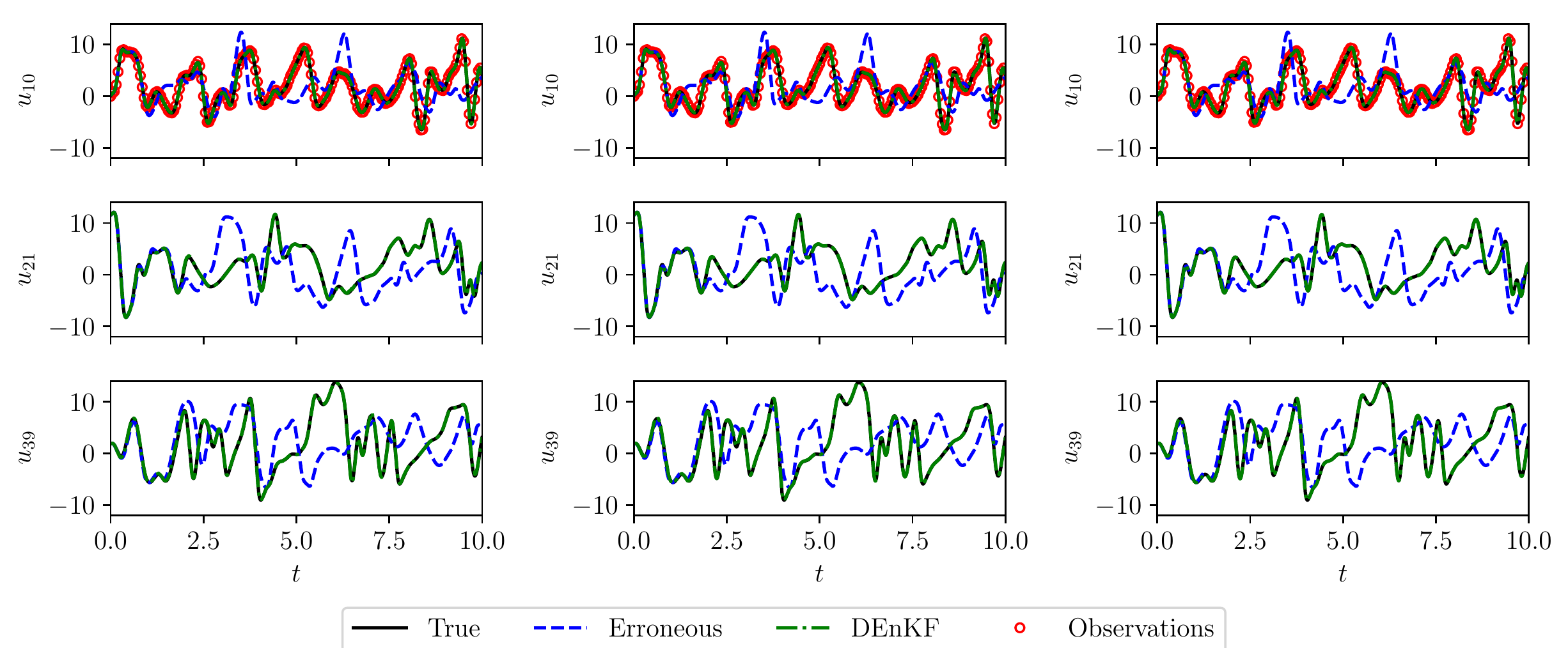}}}
\caption{Selected trajectories of the Lorenz 96 model with the analysis performed by the deterministic ensemble Kalman filter (DEnKF) with $N=40$ member ensemble using observations from $m=4$ (left), $m=8$ (middle), and  $m=20$ (right) state variables at every 10 time steps.} 
\label{fig:denkf_3}
\end{figure*}

\begin{figure*}
\centering
\mbox{\subfigure{\includegraphics[width=0.96\textwidth]{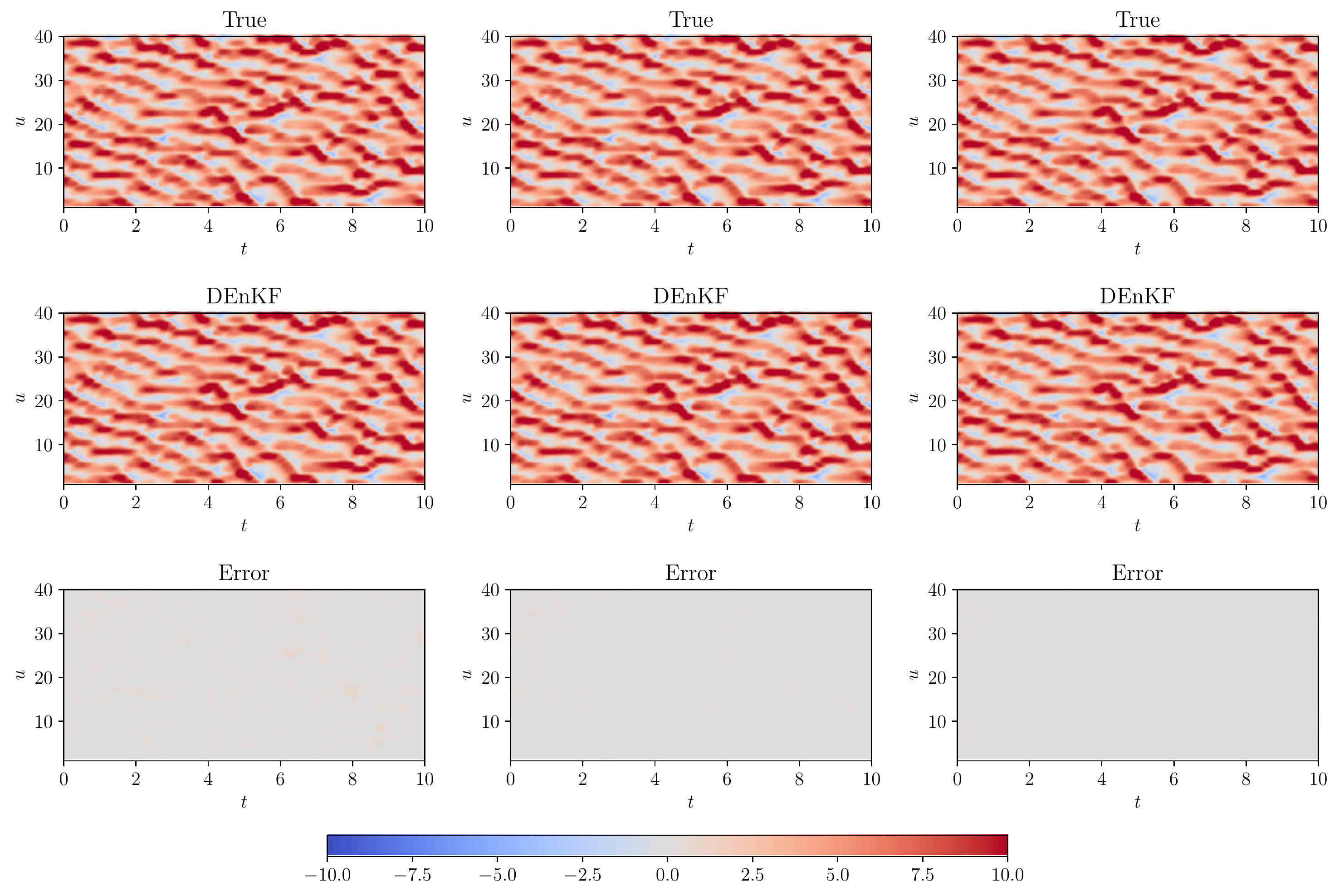}}}
\caption{Full state trajectory of the Lorenz 96 model with the analysis performed by the deterministic ensemble Kalman filter (DEnKF) with $N=40$ member ensemble using observations from $m=4$ (left), $m=8$ (middle), and  $m=20$ (right) state variables at every 10 time steps.} 
\label{fig:denkf_all}
\end{figure*}

As we observed in Figure~\ref{fig:enkf_all}, the use of virtual observations in the EnKF algorithm leads to suboptimal performance when fewer observations are used for assimilation with a small number of ensembles. The EnKF data solution converges towards a true solution with an increase in the number of ensembles. The DEnKF algorithm is the deterministic version of the EnKF algorithm where no virtual observations are used. Instead of using virtual observations, the DEnKF algorithm updates the ensemble mean with standard analysis equation and ensemble anomalies are updated separately with half the Kalman gain in the same equation \cite{sakov2008deterministic}. In Figure~\ref{fig:denkf_3}, we illustrate the time evolution of selected states for different percentages of observations used in the assimilation step. We notice that even with just four observations, the DEnKF algorithm is able to correct the erroneous states up to the final time $t=10$. From the results depicted in Figure~\ref{fig:denkf_all}, we can deduce that the DEnKF algorithm leads to better performance than the EnKF algorithm when the number of observations is smaller.  

\begin{figure*}
\centering
\mbox{\subfigure{\includegraphics[width=0.96\textwidth]{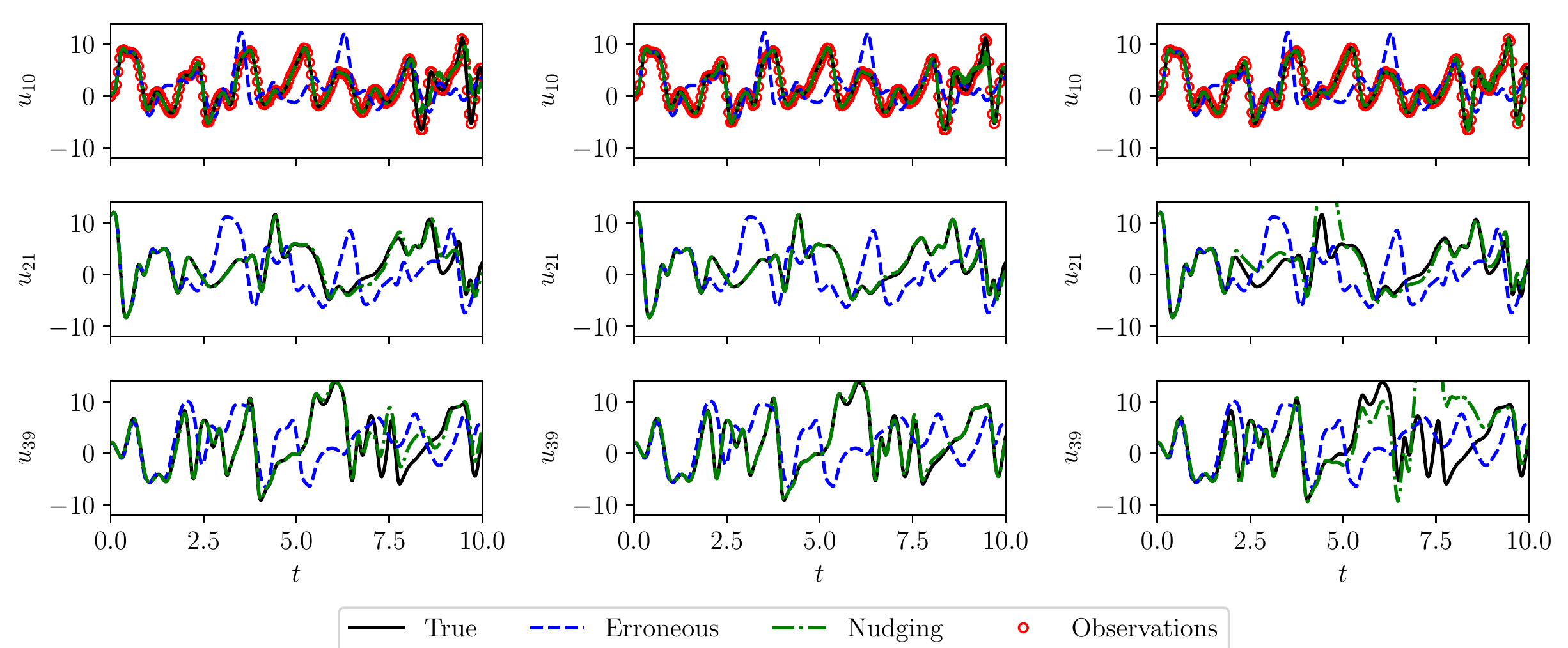}}}
\caption{Selected trajectories of the Lorenz 96 model with the analysis performed by the forward nudging with $m=20$ observations state variables at every 10 time steps for $\tau=50\Delta t$ (left), $\tau=100\Delta t$ (middle), and  $\tau=200\Delta t$ (right).} 
\label{fig:nudging_tau_3}
\end{figure*}

\begin{figure*}
\centering
\mbox{\subfigure{\includegraphics[width=0.96\textwidth]{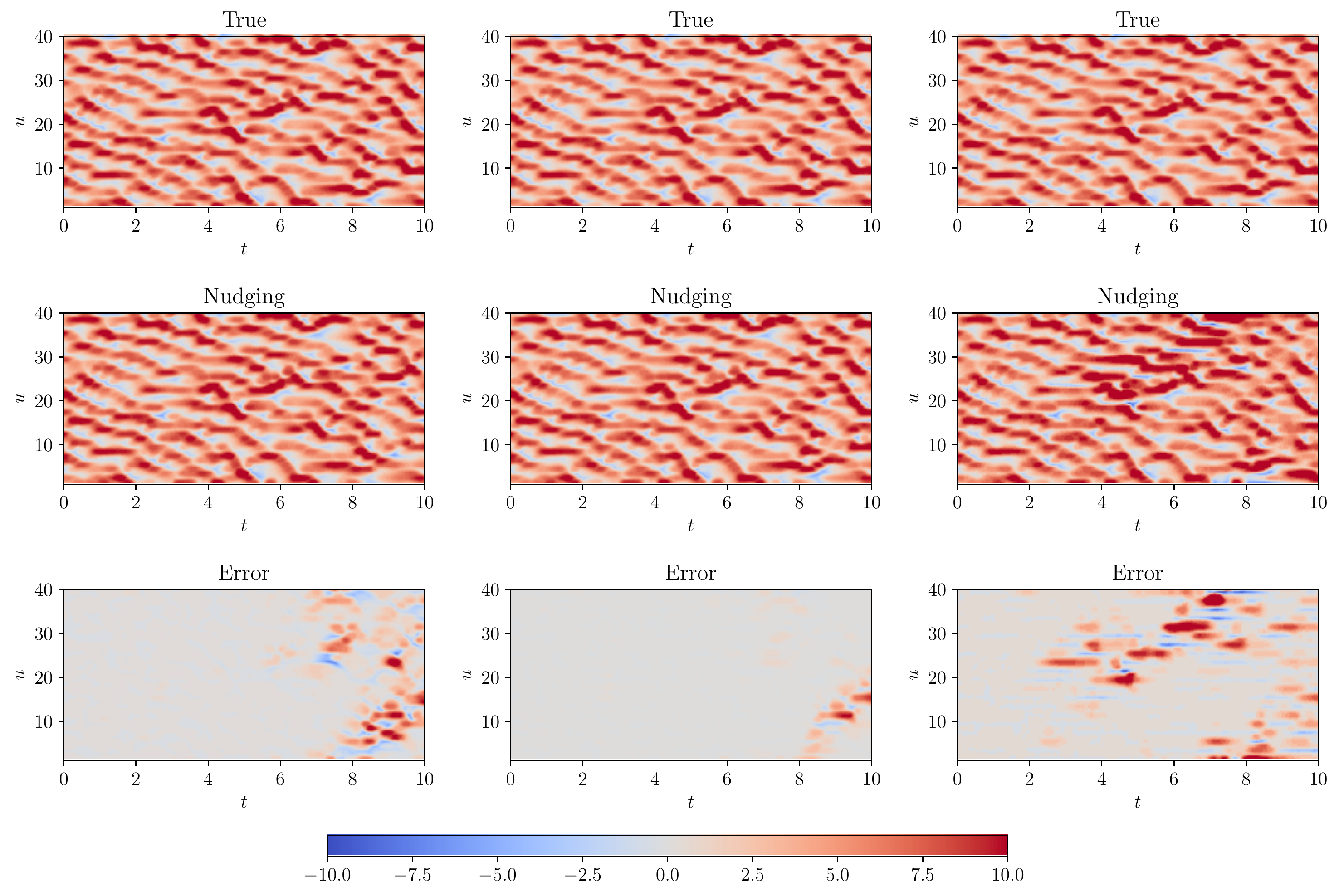}}}
\caption{Full state trajectory of the Lorenz 96 model with the analysis performed by the forward nudging with $m=20$ observations state variables at every 10 time steps for $\tau=50\Delta t$ (left), $\tau=100\Delta t$ (middle), and  $\tau=200\Delta t$ (right).} 
\label{fig:nudging_tau_all}
\end{figure*}

The nonlinear filtering methods discussed in Section~\ref{sec:kalman} are computationally expensive and are prone to curse of dimensionality with an increase in the resolution of the forward numerical model. Nudging methods, on the other hand, are computationally inexpensive and straightforward to implement. As described in Section~\ref{sec:nudging}, nudging is accomplished by adding a correction term to the dynamical model which is proportional to the difference between observations and model forecast. One of the main limitations of nudging methods is the adhoc specification of the nudging relaxation coefficient and it is not clear how to choose this coefficient to obtain an optimal solution \cite{blum2009data}. Here, we demonstrate how the choice of nudging coefficient affects the prediction when 20 observations are available for assimilation. Since the nudging coefficient represents the relaxation of time scale, we use a constant value for the nudging coefficient that is a function of the time step of the model. Also, the nudging coefficient is assumed to be constant throughout the time integration, i.e., $G_k=\tau$, where $\tau$ is a function of the time step of the model. Figure~\ref{fig:nudging_tau_3} displays the time evolution of selected states for three different values of the nudging coefficient. We notice that for a higher value of $\tau$, the nudging method is not able to correct the model forecast accurately. Figure~\ref{fig:nudging_tau_all} provides the full state trajectory of the Lorenz 96 model with different nudging coefficient matrix. We observe that the error is sufficiently low at $\tau=100 \Delta t$. Though, the prediction can be further refined by fine-tuning of the nudging coefficient matrix. 

\begin{figure*}
\centering
\mbox{\subfigure{\includegraphics[width=0.96\textwidth]{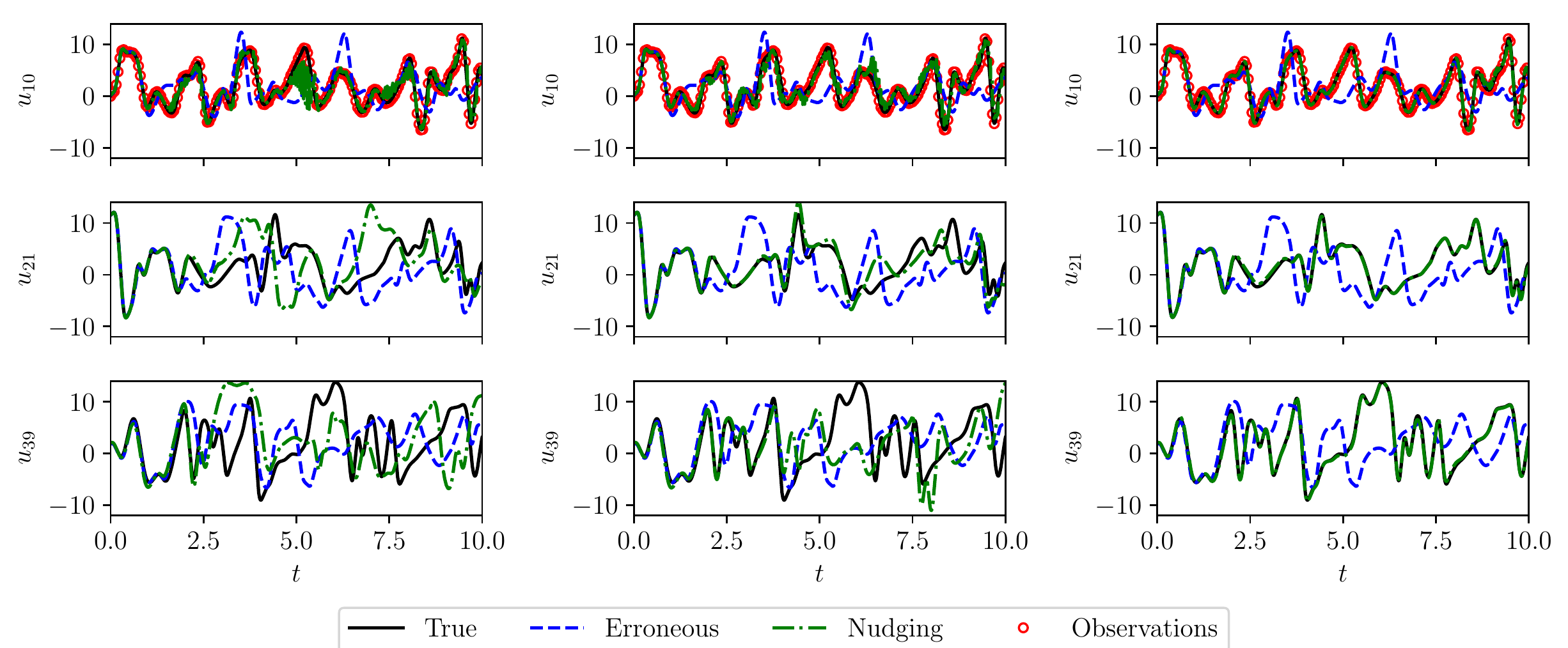}}}
\caption{Selected trajectories of the Lorenz 96 model with the analysis performed by the forward nudging with $\tau=150\Delta t$ using observations from $m=4$ (left), $m=8$ (middle), and  $m=20$ (right) state variables at every 10 time steps.} 
\label{fig:nudging_3}
\end{figure*}

\begin{figure*}
\centering
\mbox{\subfigure{\includegraphics[width=0.96\textwidth]{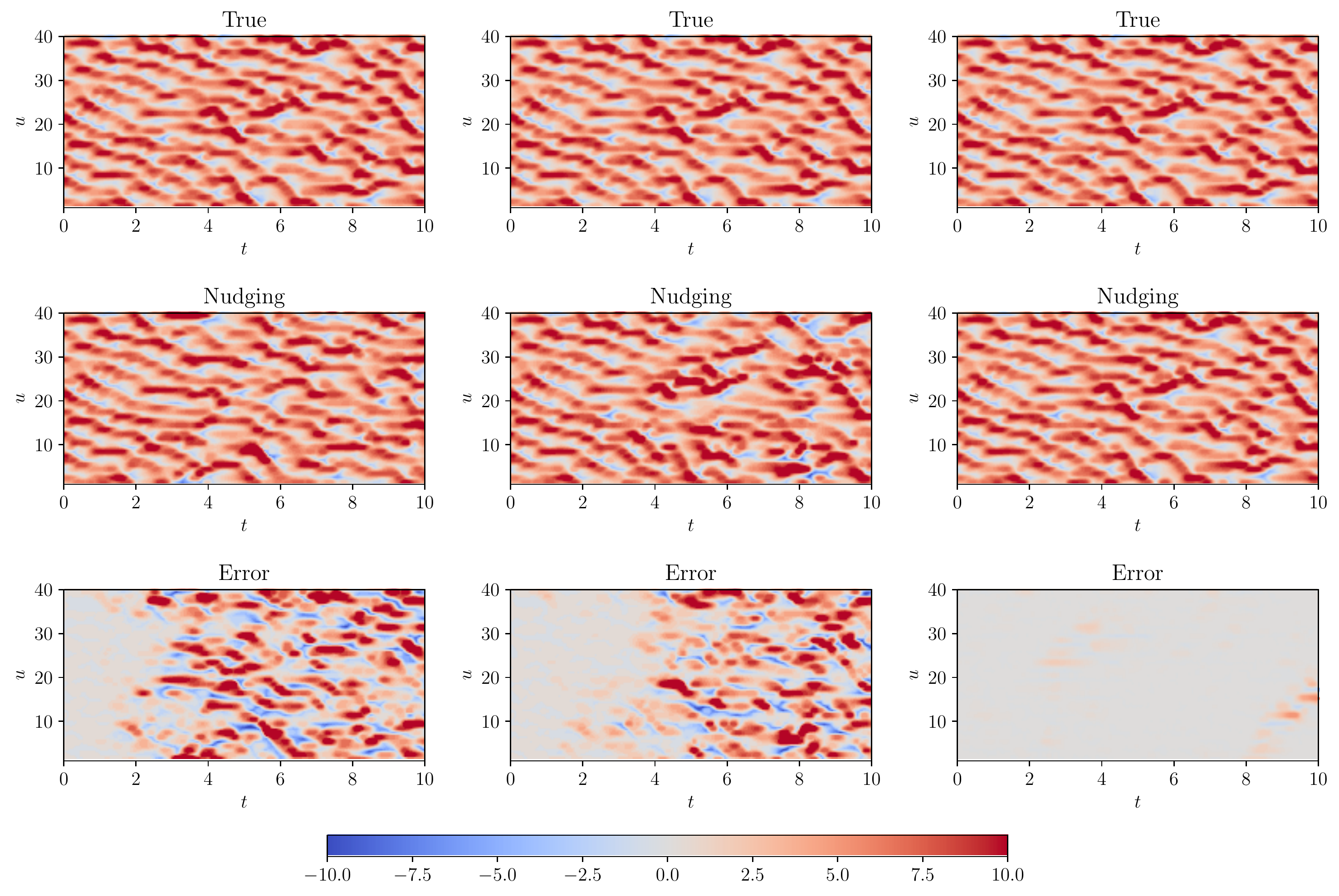}}}
\caption{Full state trajectory of the Lorenz 96 model with the analysis performed by the forward nudging with $\tau=150\Delta t$ using observations from $m=4$ (left), $m=8$ (middle), and  $m=20$ (right) state variables at every 10 time steps.} 
\label{fig:nudging_all}
\end{figure*}

Figure~\ref{fig:nudging_3} and Figure~\ref{fig:nudging_all} present the time evolution of selected states and full Lorenz 96 system for different number of observations with $\tau=150 \Delta t$. We can easily see that the prediction capability of the nudging scheme is poor when less number of observations are available for the assimilation. Indeed the performance of the nudging scheme can be improved by the optimal specification of nudging coefficient \cite{zou1992optimal}, or by using back and forth nudging algorithm \cite{auroux2005back}. However, the optimal nudging coefficient computation involves obtaining an adjoint model and solving a constrained minimization problem. Also, the back and forth nudging algorithm requires $O(10)$ iterations for convergence and the computational cost will be large for high-dimensional systems. Therefore, machine learning algorithms that are successful in finding the nonlinear mapping between two quantities can be exploited to learn the nudging dynamics.

\begin{figure*}
\centering
\mbox{\subfigure{\includegraphics[width=0.96\textwidth]{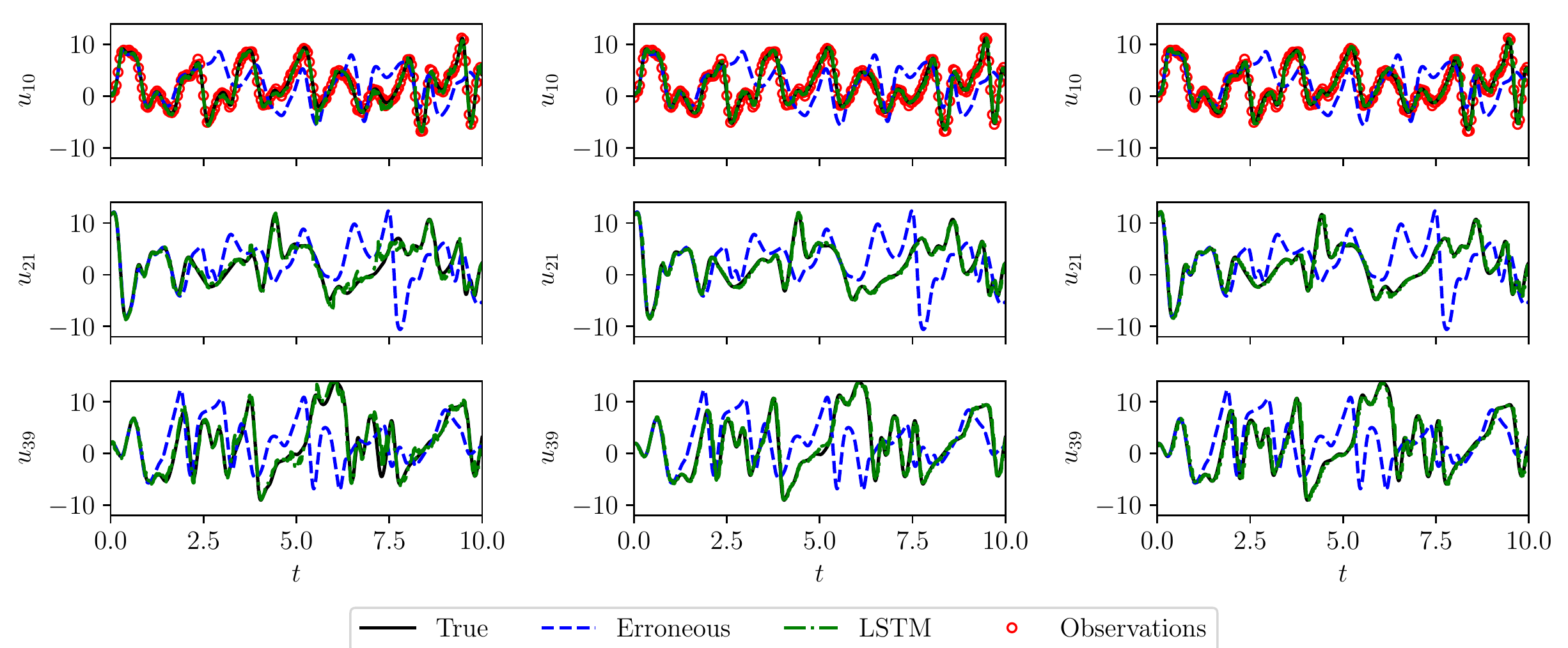}}}
\caption{Selected trajectories of the Lorenz 96 model with the analysis performed by the LSTM nudging with $N=40$ member ensemble for training using observations from $m=4$ (left), $m=8$ (middle), and  $m=20$ (right) state variables at every 10 time steps.} 
\label{fig:lstm_3}
\end{figure*}

\begin{figure*}
\centering
\mbox{\subfigure{\includegraphics[width=0.96\textwidth]{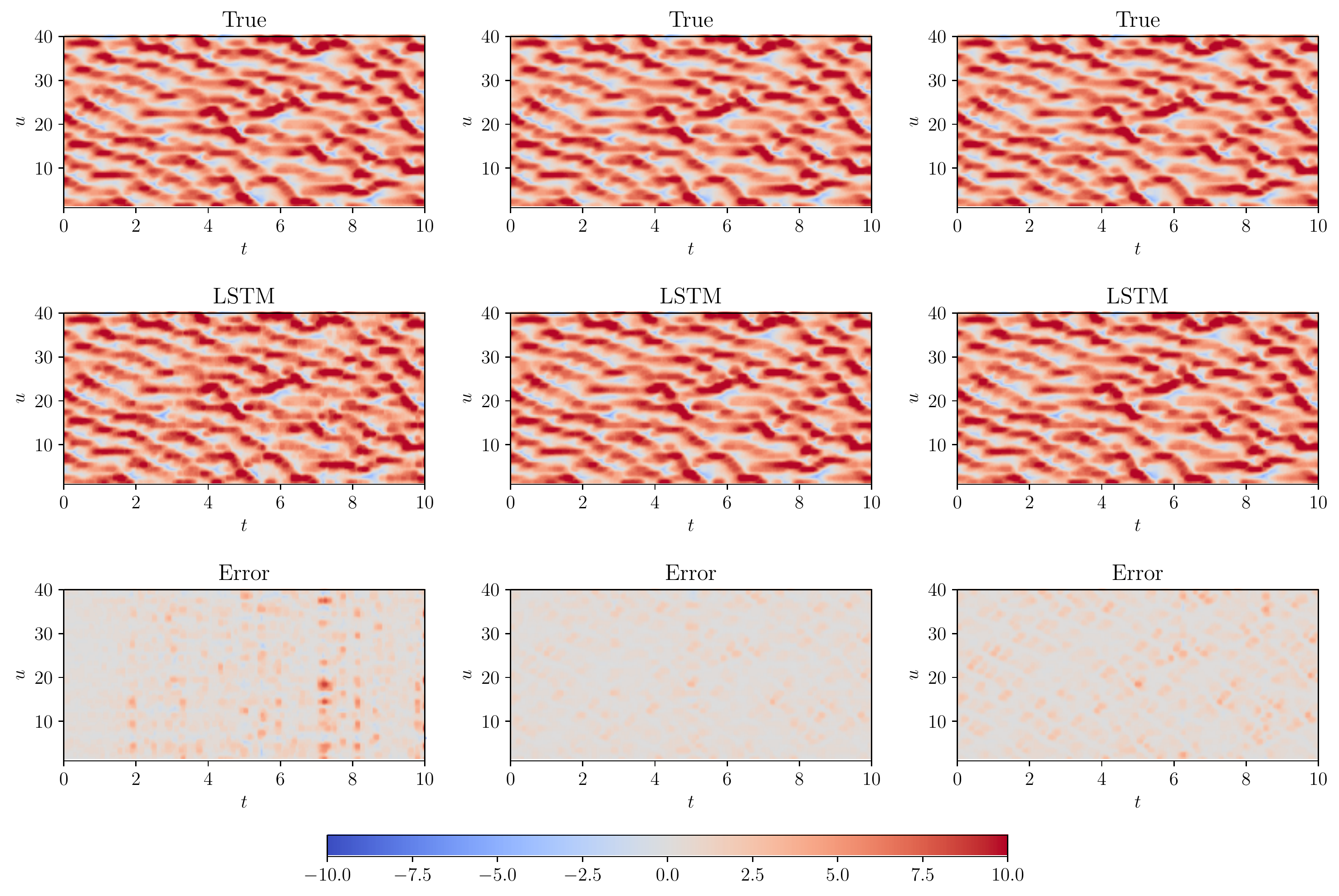}}}
\caption{Full state trajectory of the Lorenz 96 model with the analysis performed by the LSTM nudging with $N=40$ member ensemble for training using observations from $m=4$ (left), $m=8$ (middle), and  $m=20$ (right) state variables at every 10 time steps.} 
\label{fig:lstm_all}
\end{figure*}

Now, we describe the results of numerical experiments with the LSTM nudging scheme described in Section~\ref{sec:LSTM}. For the fair comparison with the EnKF and DEnKF algorithms, we use the data generated from $N=40$ perturbed initial conditions for the training of the LSTM network. These perturbed initial conditions are created by adding noise from the Gaussian distribution of zero mean and $1 \times 10^{-2}$ variance to the erroneous initial condition. The training data is obtained by integrating the model with these perturbed initial conditions from time $t=0$ to $t=10$ with $dt=5 \times 10^{-3}$ and then storing the states at all times where observations are present. Therefore, there will be 40,000 samples available for training the LSTM network. The LSTM network is trained using the procedure described in Algorithm~\ref{alg:lstmt}. We use fairly simple LSTM architecture with two hidden layers consisting of 80 LSTM cells each, and train the network for 2500 epochs. We apply the ReLU activation function and Adam optimizer for the optimization. We found that our training is not highly sensitive to neural network hyperparameters and a similar level of accuracy can be achieved with other sets of hyperparameters. Figure~\ref{fig:lstm_3} presents the time evolution of selected states for three different number of observations. We see that the LSTM network has learned the mapping from input data to the correction term and is able to produce the correct trajectory even for those states for which observations are not available. In Figure~\ref{fig:lstm_all}, we provide the full state trajectory of the Lorenz 96 model for the LSTM nudging method. We get a sufficient level of accuracy comparable to nonlinear filtering algorithms with  20\% observations. 

\begin{figure*}
\centering
\mbox{\subfigure{\includegraphics[width=0.96\textwidth]{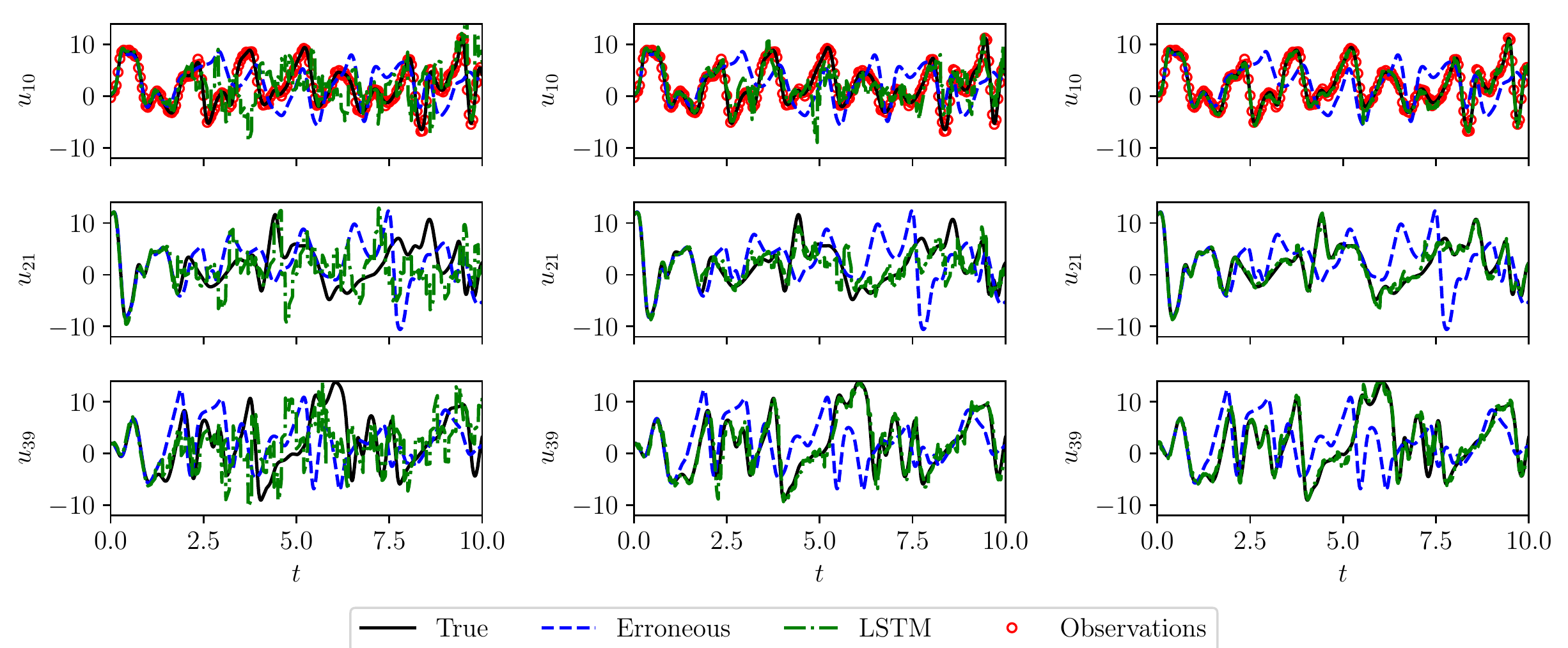}}}
\caption{Selected trajectories of the Lorenz 96 model with the analysis performed by the LSTM nudging with $N=40$ member ensemble for training using observations from $m=2$ (left), $m=3$ (middle), and  $m=4$ (right) state variables at every 10 time steps. } 
\label{fig:lstm_sparse_3}
\end{figure*}

\begin{figure*}
\centering
\mbox{\subfigure{\includegraphics[width=0.96\textwidth]{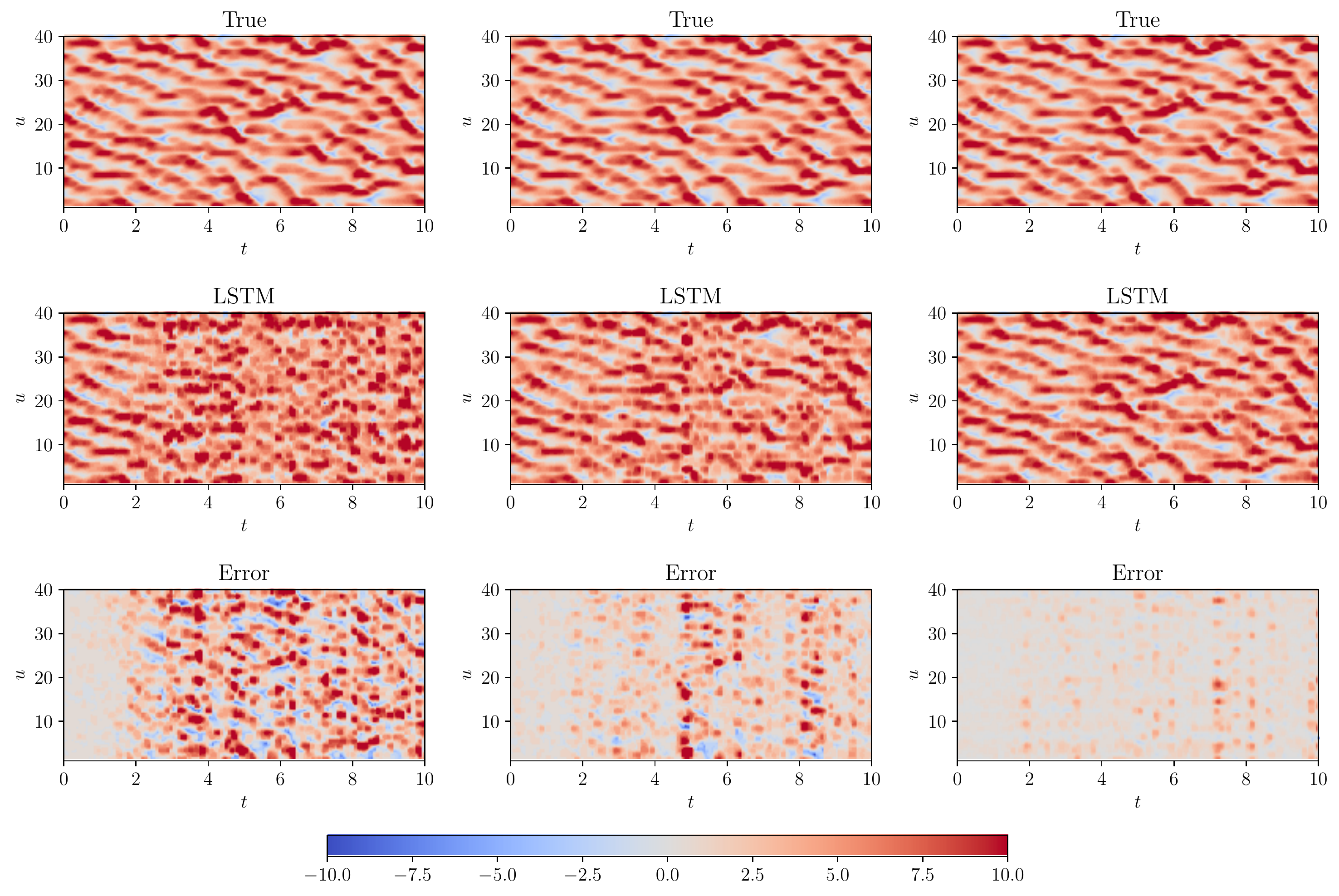}}}
\caption{Full state trajectory of the Lorenz 96 model with the analysis performed by the LSTM nudging with $N=40$ member ensemble for training using observations from $m=2$ (left), $m=3$ (middle), and  $m=4$ (right) state variables at every 10 time steps.} 
\label{fig:lstm_sparse_all}
\end{figure*}

From our analysis of numerical experiments with three sets of observations, we can conclude that the LSTM network can learn the nudging dynamics efficiently. Some of the other questions that we want to investigate in this study are; how sparse can the observations be for an accurate prediction?, and how much training data is required for training the network effectively? Figure~\ref{fig:lstm_sparse_3} displays the time evolution selected states for the LSTM nudging scheme with very sparse observations, i.e., $m=2,3,$ and $4$ and we observe a large discrepancy between true and predicted states with less than 10\% observations. Figure~\ref{fig:lstm_sparse_all} reports the full state trajectory for the Lorenz 96 model with very sparse observations. The results in Figure~\ref{fig:lstm_sparse_3} and Figure~\ref{fig:lstm_sparse_all} suggests that at least 10\% observations are necessary for producing the correct prediction with low error. We point out here that we utilized the data created from only 40 perturbed initial conditions for training, and it is well known that the performance of the neural network can be improved by training with more data. 

\begin{figure*}
\centering
\mbox{\subfigure{\includegraphics[width=0.96\textwidth]{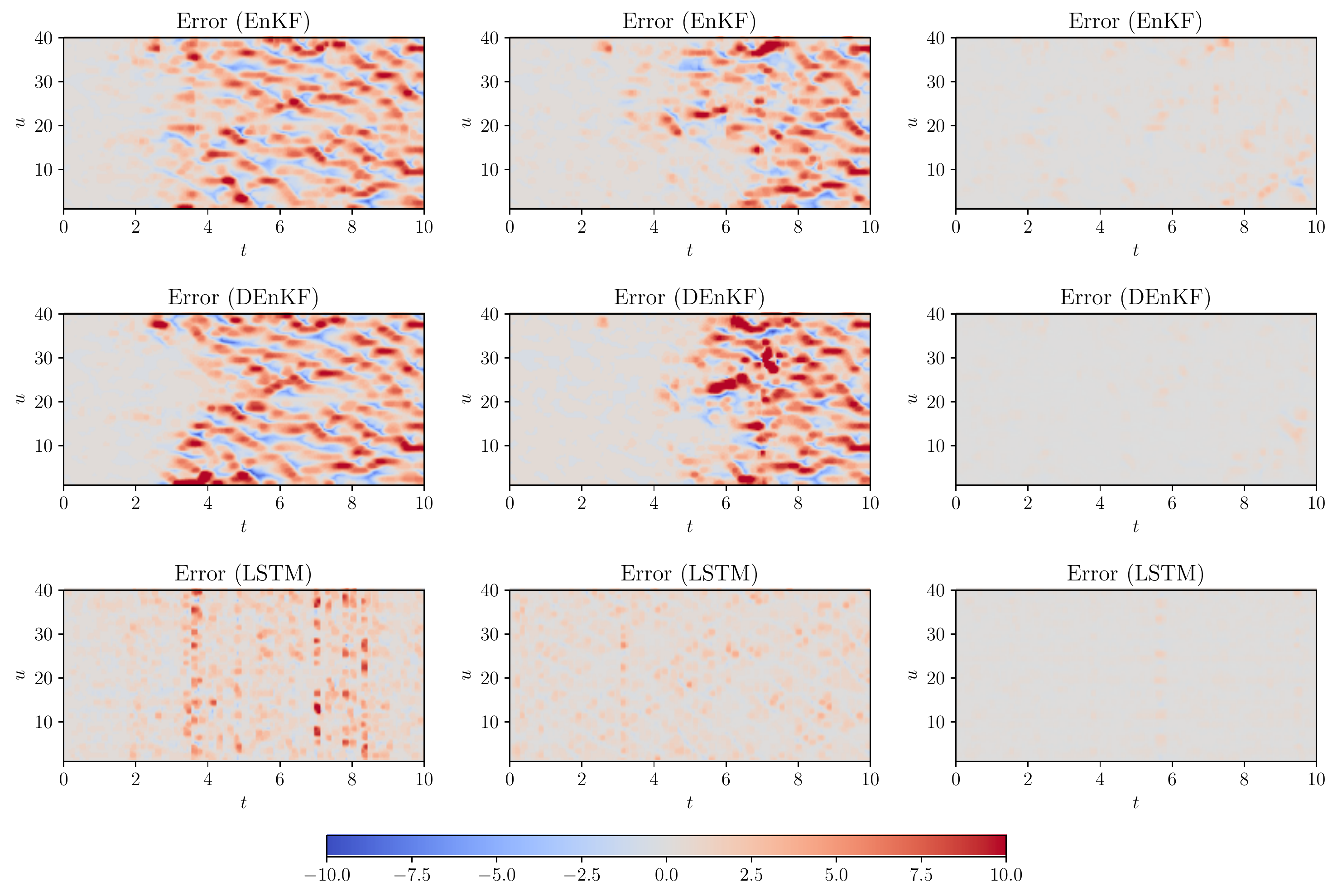}}}
\caption{Full state error of the Lorenz 96 model with the analysis performed by the ensemble Kalman filter (EnKF), deterministic ensemble Kalman filter (DEnKF), and LSTM nudging with $N=200$ member ensemble using observations from $m=2$ (left), $m=3$ (middle), and  $m=4$ (right) state variables at every 10 time steps. } 
\label{fig:error_field}
\end{figure*}

\begin{figure*}
\centering
\mbox{\subfigure{\includegraphics[width=0.96\textwidth]{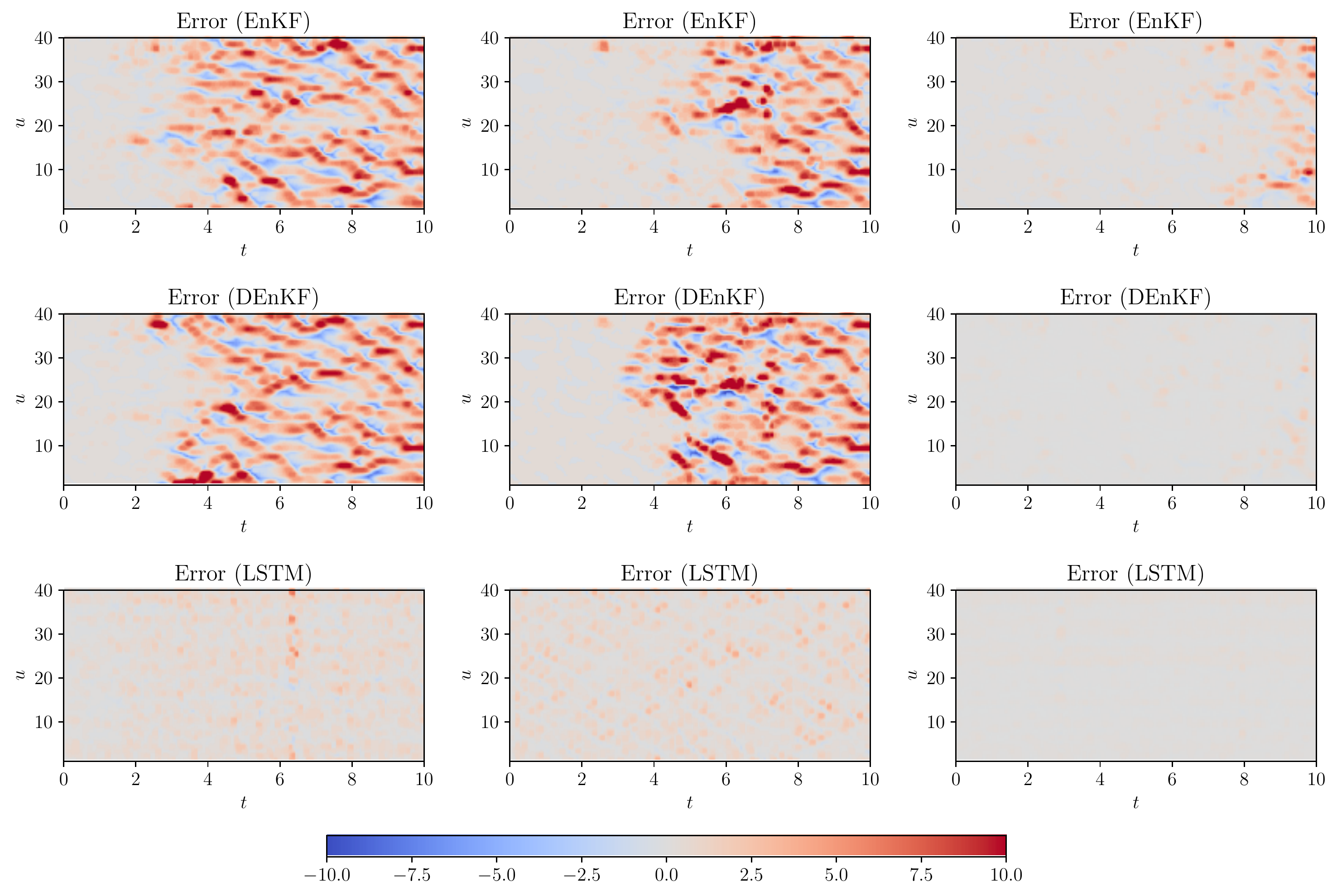}}}
\caption{Full state error of the Lorenz 96 model with the analysis performed by the ensemble Kalman filter (EnKF), deterministic ensemble Kalman filter (DEnKF), and LSTM nudging with $N=400$ member ensemble using observations from $m=2$ (left), $m=3$ (middle), and  $m=4$ (right) state variables at every 10 time steps.} 
\label{fig:error_field400}
\end{figure*}

In Figure~\ref{fig:error_field} and Figure~\ref{fig:error_field400}, we illustrate the improvement in prediction for highly sparse observations as the amount of data employed for training the LSTM network is increased. We show only the error plot (the difference between true and predicted states) for the conciseness. We can easily observe that the error is large for the EnKF and DEnKF algorithms compared to the LSTM nudging scheme when only two or three observations are available for assimilation. When four observations are present, we see a similar level of accuracy for EnKF, DEnKF, and LSTM nudging method. If we compare the error in Figure~\ref{fig:error_field} and Figure~\ref{fig:error_field400}, there is an improvement in the prediction as we increase the training data. The results presented in Figure~\ref{fig:error_field} and Figure~\ref{fig:error_field400} are obtained by utilizing $N=200$ and $N=400$ ensemble members for the EnKF and DEnKF algorithms. The same number of perturbed initial conditions are also used for training the LSTM network. Therefore, in terms of computational cost, all three methods can be considered equivalent because the same number of forward numerical models are integrated from initial time to final time for all three methods. In terms of the storage, the LSTM nudging is more demanding as it requires the storage of full state for all training sets (i.e., perturbed initial condition) at all observation points for the training. However, there is no need to store the solution of all ensemble members in the EnKF and DEnKF algorithm. This limitation can be addressed by transfer learning, where the weights and biases of the neural network are updated by training its last few layers with new data. Therefore, training the LSTM network for the first time is a computationally intensive task and the LSTM network can be retrained as new observations become available. 

\begin{figure}
\centering
\mbox{\subfigure{\includegraphics[width=0.4\textwidth]{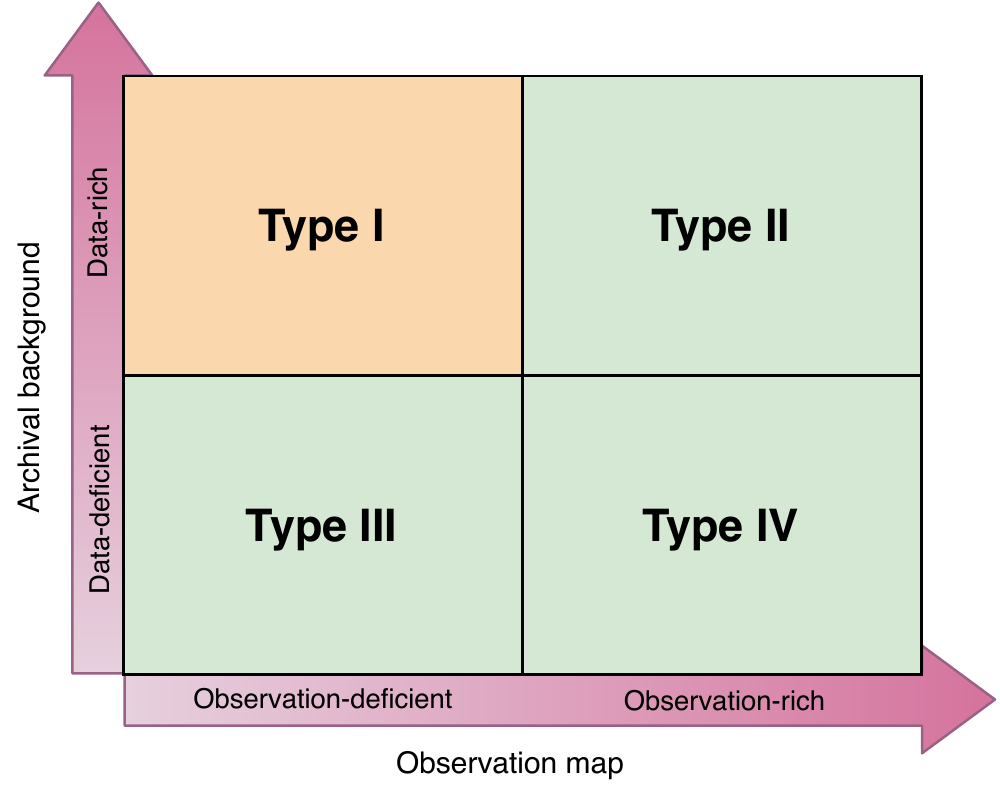}}}
\caption{Segregation of problems encountered in data-assimilation based on the observations and archival/ensemble background data. LSTM nudging method is particularly suitable where there is a rich amount archival or background information available for training the network and observations are sparse.} 
\label{fig:da_lstm_data}
\end{figure}

\section{Conclusions}
\label{sec:conc}
In the present study, we introduced the LSTM nudging scheme that learns the nudging dynamics from the full state of the system and partial observations. We illustrate the approach for the Lorenz 96 system and compare its performance against extended Kalman filter (EKF), ensemble Kalman filter (EnKF), and deterministic ensemble Kalman filter (DEnKF) approaches. We consider different aspects of the LSTM nudging scheme such as sparsity in observations, and the amount of available data for training the LSTM network. We successfully demonstrate that the LSTM network can be trained to learn the nudging dynamics with extremely sparse observations provided there is a large amount of training data. In terms of computational overhead, training the neural network is the most demanding task. However, this is a one time task and future observations can be incorporated by retraining the neural network with transfer learning at a much less computational cost.

The results of our numerical experiments with the LSTM nudging scheme indicate its potential benefit of assimilation from very sparse observations. Another benefit is that there are no matrix computational operations such as Kalman gain calculation. One of the important caveats of the LSTM nudging scheme is that the neural networks are data-hungry and hence a large amount of archival or background data will be necessary to train the neural network. The suitability of LSTM nudging scheme for DA problems is summarized in Figure~\ref{fig:da_lstm_data}, where the DA problems are classified based on the sparsity of observations and amount of archival background information. The LSTM nudging scheme is well suited for problems where observations are very sparse and there is the availability of archival background information (i.e., type I problems). Another limitation is that the training procedure in the present form will not be feasible for very high-dimensional systems. One of the solutions to address this constraint is to utilize reduced order modeling (ROM) approaches for dimensionality reduction and recently, machine learning methods are found to give accurate, stable, and robust ROMs for physical systems. Since the LSTM nudging scheme is flexible, we foresee that this approach can be extended to large scale systems by blending it with ROM approaches. One more reservation of the LSTM nudging method is that it does not predict the uncertainty in analyzed states.

We re-emphasize here that the significance of the proposed LSTM nudging method on the prototype model does not mean that they can be directly extended to higher-dimensional and more complex problems. In this work, we assumed that the model is perfect and the noise is Gaussian, which is a very idealized condition. In actual scenarios, real weather forecast models are approximate and contain a lot of parameterizations for subgrid scale processes. Therefore, one can look at the results of numerical experiments presented in this study as the early findings and substantial future work is required for the demonstration of the proposed method in a realistic situation. As a part of future studies, we plan to illustrate the LSTM nudging method for a two-dimensional quasi-geostrophic model with an application of convolutional autoencoder for dimensionality reduction. Neural networks have also been shown to be capable of discovering hidden information about the physical processes embedded in the data \cite{raissi2018hidden,pawar2020data} and we will integrate these methods with the LSTM nudging scheme for imperfect models.

\section*{Acknowledgement}
This material is based upon work supported by the U.S. Department of Energy, Office of Science, Office of Advanced Scientific Computing Research under Award Number DE-SC0019290. O.S. gratefully acknowledges their support. 

Disclaimer. This report was prepared as an account of work sponsored by an agency of the United States Government. Neither the United States Government nor any agency thereof, nor any of their employees, makes any warranty, express or implied, or assumes any legal liability or responsibility for the accuracy, completeness, or usefulness of any information, apparatus, product, or process disclosed, or represents that its use would not infringe privately owned rights. Reference herein to any specific commercial product, process, or service by trade name, trademark, manufacturer, or otherwise does not necessarily constitute or imply its endorsement, recommendation, or favoring by the United States Government or any agency thereof. The views and opinions of authors expressed herein do not necessarily state or reflect those of the United States Government or any agency thereof.

\appendix
\section{Jacobian of the model and observation matrix}

We apply a fourth-order Runge-Kutta (RK4) numerical scheme for temporal integration of the Lorenz 96 model and it can be written as follow
 \begin{equation} \label{eq:rk4}
     \mathbf{u}^{k+1} = \mathbf{u}^{k} + \frac{\Delta t}{6}(\mathbf{g}_1 + 2 \mathbf{g}_2 + 2 \mathbf{g}_3 + \mathbf{g}_4),
 \end{equation}
where
\begin{align}
    \mathbf{g}_1 &= \mathbf{f}(\mathbf{u}^{k}), \\
    \mathbf{g}_2 &= \mathbf{f}(\mathbf{u}^{k} +\frac{\Delta t}{2}\cdot \mathbf{g}_1), \\
    \mathbf{g}_3 &= \mathbf{f}(\mathbf{u}^{k} +\frac{\Delta t}{2}\cdot \mathbf{g}_2), \\
    \mathbf{g}_4 &= \mathbf{f}(\mathbf{u}^{k} +\Delta t \cdot \mathbf{g}_3).
\end{align}
The function $\mathbf{f}$ is the right hand ride of the Lorenz 96 model and in the discrete form it can be written as 
\begin{equation}
    f_i = u_{i-1} (u_{i+1} - u_{i-2}) - u_i + F.
\end{equation}
The Jacobian of the function $\mathbf{f}$ is defined as below
\begin{equation}
    \mathbf{J} = \frac{\partial \mathbf{f}}{\partial \mathbf{u}} = \bigg[\frac{\partial f_i}{\partial u_j} \bigg] \quad \text{for}~ 1 \le i,j \le n.
\end{equation}
The Jacobian $\mathbf{J}$ will be a $\mathbb{R}^{n \times n}$ matrix. The Jacobian of the model $\mathbf{D_M} \in \mathbb{R}^{n \times n}$ can be computed by applying the chain rule to Equation~\ref{eq:rk4} and is given below
\begin{equation}
    \mathbf{D_M} = \mathbf{I} + \Delta t \cdot \mathbf{J} + \frac{1}{2} {\Delta t}^2 \cdot \mathbf{J}^2 + \frac{1}{6} {\Delta t}^3 \cdot \mathbf{J}^3 + \frac{1}{24} {\Delta t}^4 \cdot \mathbf{J}^4,
\end{equation}
where $\mathbf{I} \in \mathbb{R}^{n \times n}$ is an identity matrix.

The Jacobian of observation is denoted by $\mathbf{D_h} \in \mathbb{R}^{m \times n}$ and is computed as shown below
\begin{equation}
    \mathbf{D_h} = \bigg[\frac{\partial h_i}{\partial u_j} \bigg],
\end{equation}
where $1 \le i \le m$ and $1 \le j \le n$. Since we use linear observations, $\mathbf{D_h}$ will be a constant sparse matrix. Each row of the matrix $\mathbf{D_h}$ will consist of all zeros except for the corresponding observation location, where it will have the value of one.

\bibliographystyle{unsrt} 
\bibliography{ref}   

\end{document}